\documentclass[english]{iopart}

\usepackage{graphicx}
\usepackage{url}
\usepackage{amssymb}
\usepackage{epstopdf}
\begin{document}

\title[$\, \tilde{\chi}^{(5)}$ ODE in exact arithmetic]
{Square lattice Ising model $\, \tilde{\chi}^{(5)}$ ODE  in exact arithmetic}

\author{B. Nickel$^\P$, I.~Jensen$^\ddag$,
S. Boukraa$^\dag$,  A. J. Guttmann$^\ddag$, S. Hassani$^\S$,
J.-M. Maillard$^{||}$ and N. Zenine$^\S$}

\address{$^\P$ Department of Physics, University of Guelph, Guelph, 
Ontario N1G 2W1, Canada} 
\address{$\ddag$ ARC Centre of Excellence for 
Mathematics and Statistics of Complex Systems 
Department of Mathematics and Statistics,
The University of Melbourne, Victoria 3010, Australia}
\address{\dag LPTHIRM and D\'epartement d'A{\'e}ronautique,
 Universit\'e de Blida, Algeria}
\address{\S  Centre de Recherche Nucl\'eaire d'Alger, 
2 Bd. Frantz Fanon, BP 399, 16000 Alger, Algeria}
\address{$||$ LPTMC, UMR 7600 CNRS, 
Universit\'e de Paris, Tour 24,
 4\`eme \'etage, case 121, 
 4 Place Jussieu, 75252 Paris Cedex 05, France} 

\ead{bgn@physics.uoguelph.ca, I.Jensen@ms.unimelb.edu.au, boukraa@mail.univ-blida.dz,
tonyg@ms.unimelb.edu.au, maillard@lptmc.jussieu.fr,  
njzenine@yahoo.com}

\begin{abstract}

We obtain in exact arithmetic the order 24 linear differential operator $\, L_{24}$ and 
right hand side $E^{(5)}$ of the inhomogeneous equation $\, L_{24}(\Phi^{(5)}) = E^{(5)}$,
where $\, \Phi^{(5)}\, = \,\,$ $ \tilde{\chi}^{(5)}\, -\tilde{\chi}^{(3)}/2\, +\, \tilde{\chi}^{(1)}/120$ is
a linear combination of $\, n$-particle contributions to the susceptibility of the square
lattice Ising model.  In Bostan \etal  (J. Phys. A: Math. Theor. {\bf 42}, 275209 (2009)) the 
operator $\, L_{24}$ (modulo a prime) was shown to factorize into 
$\, L_{12}^{(\rm left)} \cdot \, L_{12}^{(\rm right)}$; here we prove that no further factorization of the
order 12 operator $\, L_{12}^{(\rm left)}$ is possible. We use the exact ODE to obtain
the behaviour of $ \tilde{\chi}^{(5)}$ at the ferromagnetic critical point and to obtain a 
limited number of analytic continuations of $ \tilde{\chi}^{(5)}$ beyond the principal disk 
defined by its high temperature series.  Contrary to a speculation in Boukraa \etal
(J. Phys. A: Math. Theor. {\bf 41} 455202 (2008)), we find that $ \tilde{\chi}^{(5)}$ 
is singular at $\, w=\, 1/2$ on an infinite number of branches.
\end{abstract}

\vskip .5cm


\noindent {\bf AMS Classification scheme numbers}: 34M55, 
47E05, 81Qxx, 32G34, 34Lxx, 34Mxx, 14Kxx 


\newpage

\section{Introduction}
\label{int}
\vskip .1cm

The story of the zero-field susceptibility $\chi$ of the two-dimensional Ising model is a landmark 
saga of mathematical physics. A recent review of the highlights can be found in \cite{M10}. 
While a closed form expression for the susceptibility still eludes us, we possess an 
enormous amount of exact or extremely precise numerical information. This largely 
derives from two complementary approaches. One approach involves studying the 
series expansion of the susceptibility. Since the work of Orrick \etal \cite{ONGP01} 
we have had available a polynomial time algorithm, now of complexity O$(N^4)$ 
for a series of length $N$ terms. From the point of view of an algebraic combinatorialist, 
this comprises a solution, and many questions about the asymptotics, and about the 
scaling functions have been answered by analysis of the very long series we now 
have available--currently more than two thousand terms in length.

The difficulty in proceeding further with this approach is that we have no idea what the 
underlying closed-form solution looks like, except that it is known, or more precisely 
universally believed, to be non-holonomic. The alternative approach is to express the 
susceptibility as a form-factor expansion. This approach was initiated more than 30 
years ago by Wu \etal \cite{WMTB76}. In this representation, the susceptibility is written
\begin{equation}
k_BT\, \chi = (1-t)^{1/4}\sum_{n \ge 1} {\tilde \chi}^{(2n+1)}
\end{equation}
for $T > T_c,$ where $t = \sinh^4 (J/k_B T).$
For $T < T_c$ a similar expression with even superscripts prevails. The advantage of
the form-factor approach is that  each term in the sum {\it is} holonomic. This means that, 
with sufficient computational resources, and sufficient ingenuity, each term can be found. 
To date, the first  six  terms have been found, in the sense that their defining ODE has 
been obtained, either totally, or {\it modulo} a prime.  We also have precise integral 
representations for the form-factor terms, and a Landau analysis of the integrands can 
provide information as to the distribution of singularities in the complex plane. Indeed, 
it was just such a study by Nickel \cite{nickel-99, nickel-00} that gave convincing evidence 
of a natural boundary in the {\it total} susceptibility, thus supporting an earlier but weaker 
argument of Guttmann and Enting \cite{GE96} that the total susceptibility was non-holonomic.

While an exact solution for the Ising model susceptibility may be impossible and is certainly 
beyond reach at present, one might hope to obtain a complete picture of the singularities of 
the susceptibility.  Indeed, this more limited goal has been the main motivation of the recent 
studies of the individual form-factor terms.  For this the ODE and Landau analysis approaches 
are complementary; the Landau analysis provides necessary but not sufficient conditions \cite{Hwa}
while the ODE, even if only in mod prime representation, can show which Landau singularities 
are to be excluded.  The most detailed study in this regard is that of the five particle contribution 
$\tilde{\chi}^{(5)}$ to the susceptibility initiated by Boukraa \etal \cite{experim} and followed by 
Bostan \etal \cite{High}.  The present paper is an attempt to address a number of issues 
left unresolved in these papers.

A brief summary of the parts of \cite{experim} and \cite{High} relevant here is as follows.  
In Boukraa \etal \cite{experim} series in   $\, w$ modulo a prime to 10000 terms were given
and shown to be adequate to find the order 33 Fuchsian differential 
equation\footnote[2]{The notation used here is that in~\cite{experim}: variables 
$\, w$ and $\, s$ are useful for high temperature expansions with $\, w\, =\, s/2/(1+s^2)$.  
The high temperature $\, \chi^{(2n+1)}$ and $\, \tilde{\chi}^{(2n+1)}$ are related by 
$ \, s \cdot \chi^{(2n+1)}\, =\, \, (1-s^4)^{1/4} \cdot \tilde{\chi}^{(2n+1)}$ so that  
$\, \tilde{\chi}^{(2n+1)}$ has the ``simpler'' divergence 
$\propto \,  1/(1-4w) \, \propto \,  1/(1-s)^2$ at
the ferromagnetic critical point.}, modulo a prime, $L_{33}(\tilde{\chi}^{(5)})\, =\,  0$. 
Subsequently in Bostan \etal~\cite{High},  the complexity of this ODE was shown to 
be reducible to an inhomogeneous equation
\begin{eqnarray}
\label{eq:L24}
L_{24} (\Phi^{(5)})  \,\, = \, \, \,\, E^{(5)},  
\end{eqnarray}
where $\, \Phi^{(5)}$ is the linear combination of 5, 3 and 1-particle contributions
\begin{eqnarray}
\label{eq:Phi5}
\, \Phi^{(5)}\, = \,\, \tilde{\chi}^{(5)}\, -\tilde{\chi}^{(3)}/2\, +\, \tilde{\chi}^{(1)}/120.
\end{eqnarray}
The right hand side of (\ref{eq:L24}) which satisfies $\,L_5(E^{(5)}) = \, 0$ is of the form
\begin{eqnarray}
\label{eq:E5}
&&E^{(5)}\,  = \, \, 
  w \cdot [(1-16\, w^2)^3 \cdot P_{4,0} \cdot K^4 \, 
+(1-16\, w^2)^2\cdot P_{3,1} \cdot K^3\, E \,
\nonumber \\
&&\quad \quad  +(1-16\, w^2)\cdot P_{2,2} \cdot K^2 \cdot  E^2
         \,  + \, P_{1,3}\cdot K\cdot  E^3 \,
\nonumber \\
&&\quad  \quad  +P_{0,4}\cdot E^4]/(1+4w)^6/(1-16\, w^2)^\kappa,
\end{eqnarray}
where $\, K=\, K(4 \, w)$ and $\, E=\, E(4\, w)$ are complete elliptic integrals 
and the $\, P_{i,j}=\, P_{i,j}(w)$ are polynomials.  The 
degree of the polynomials and the denominator power 
$\, \kappa$ in (\ref{eq:E5}) depend on the representation of
$\, L_{24}$ in (\ref{eq:L24}). In the case that $\, L_{24}$ is 
minimum order 24, $L_{24}$ is of degree 888, while the $\, P_{i,j}$ are 
then of degree (at most) 904 with $\, \kappa\, =\,8$.  In~\cite{High}
results were only reported for a non-minimum order representation 
modulo a prime.  Furthermore it was shown that $\, L_{24}$ could be 
factored into $\, L_{12}^{(\rm left)}\, L_{12}^{(\rm right)}$ with 
$\, L_{12}^{(\rm right)}$ being reducible into several smaller factors, 
all but one known in exact arithmetic.  The question of whether 
$\, L_{12}^{(\rm left)}$ could be factored was left unresolved.

The results reported in~\cite{experim} and~\cite{High} are an impressive 
example of what can be obtained by calculation modulo a prime.  
However there are limitations. Knowing $\, L_{24}$ and $\, E^{(5)}$ modulo 
a prime enables one to deduce only possible local singularities 
of $\tilde{\chi}^{(5)}$  and not its global behaviour.  For example, one 
cannot determine the amplitudes of the singularities in $\tilde{\chi}^{(5)}$  
at the ferromagnetic point at $\, s\, =\, 1$.  The leading and first correction 
term amplitudes were estimated in~\cite{experim} but this was based on 
an analysis of the 2000 term exact integer series that had been obtained 
from multiple modulo prime series by the Chinese remainder theorem.

In the present paper we determine the minimum order $\, L_{24}$ and $E^{(5)}$ 
in exact arithmetic. The results can be found on the website~\cite{http}.  We have 
also constructed the minimum order exact $L_{29}$ defined by 
$\, L_{29}(\Phi^{(5)})\, =\,\, 0$ from these results.  We have not actually needed 
this operator but provide it nevertheless for those who might find it of interest  
and in addition we also report in~\cite{http} the exact integer coefficients of 
$\tilde{\chi}^{(5)}$ to 8000 terms that we generated directly\footnote[3]{For the 
required $\tilde{\chi}^{(3)}$ see \ref{ferro} and references therein.} 
from (\ref{eq:L24}).  The global information provided by (\ref{eq:L24}) enables 
us to prove that $\, L_{12}^{(\rm left)}$ defined by 
$\, L_{24}= \, $  $\, L_{12}^{(\rm left)}\, L_{12}^{(\rm right)}$ cannot be factored.  
We confirm the 500 digit amplitude of the leading ferromagnetic singularity of 
$\tilde{\chi}^{(5)}$ reported by Bailey \etal~\cite{Bailey} and also some 
$|s|=\, 1$ circle singularity amplitudes derived by Nickel~\cite{nickel-99,nickel-00}.

An important question left unresolved in~\cite{experim}
was whether the singularity of the ODE for $\tilde{\chi}^{(5)}$
at the zero of the head polynomial at  $\, w =\, 1/2$  was in fact a 
singularity of $\tilde{\chi}^{(5)}$ on some branch of 
the function.  Appendix D in~\cite{experim} provided an
example for which not all the zeros of the head
polynomial of the ODE\footnote[1]{We explicitly
exclude zeros associated with apparent singularities.} 
satisfied by an integral were Landau
singularities of the integral.  However, the integral
for  $\tilde{\chi}^{(5)}$ was deemed too complicated
in~\cite{experim} to perform a complete Landau 
singularity analysis leaving only the conclusion 
that $\, w =\, 1/2$ was very likely not a Landau singularity
of  $\tilde{\chi}^{(5)}$.  Here we perform some 
analytic continuation of $\tilde{\chi}^{(5)}$
beyond the principal disk $ |s|\leq \, 1$ and 
onto other branches. Our exploration, while not exhaustive, is complete enough 
to show $\tilde{\chi}^{(5)}$ has the singular behaviour
($1-2w)^{7/2}$ with non-vanishing amplitude on an infinite number of branches. 
We have not made any progress in identifying the Landau integrand 
singularities that give rise to this behaviour.

\vskip .3cm

In section \ref{ode} we describe briefly how we can combine results 
reported in~\cite{High} with multiple modulo prime $\, \Phi^{(5)}$ series 
of length at most 4000 terms to obtain the exact $\, L_{24}$ and $E^{(5)}$ 
in (\ref{eq:L24}).  This allows us to determine the series expansion of 
$\tilde{\chi}^{(5)}$ at the ferromagnetic point by numerical matching of 
solutions. The results are reported in \ref{ferro}.  Section \ref{proof} is the 
outline of the proof that $\, L_{12}^{(\rm left)}$ defined by 
$\, L_{24} = \, $  $\, L_{12}^{(\rm left)}\cdot L_{12}^{(\rm right)}$ cannot be 
factored.  Finally in section \ref{analyt} we describe the analytic continuation 
of $\, \Phi^{(5)}$   we have performed to obtain information on the behaviour
of $\,\tilde{\chi}^{(5)}$ at $\, w\, =\, 1/2$.

\section{The ODE for $\, \Phi^{(5)}$ in exact arithmetic}
\label{ode}
\vskip .1cm

We are looking for the unique minimum order $\, L_{24}$
and associated $\, E^{(5)}$  satisfying (\ref{eq:L24})
for a given $\, \Phi^{(5)}$.  This is to be done modulo a prime 
for enough primes\footnote[5]{ We have found that
about 90 primes $p<\, 2\, ^{15}$ are sufficient.} that the exact integer
$\, L_{24}$ and $\, E^{(5)}$ can be reconstructed
by the Chinese remainder theorem.  To generate the required
$\, \Phi^{(5)}$ in (\ref{eq:Phi5}) directly from the defining $\, \tilde{\chi}^{(n)}$ 
integrals is impractical as the minimum order $\, L_{24}$ is of degree 888
implying $\, 25 \times 889 \,  = \, \,  22225$ unknown
coefficients in $\, L_{24}$.  Added to this are the five
polynomials of degree 904 in $\, E^{(5)}$, i.e. another
$\, 5 \times 905\, = \, 4525$ terms.  Finding all coefficients
is a straightforward linear algebra problem but
requires that we have each $\, \Phi^{(5)}$ modulo prime
series to about 26800 terms\footnote[2]{An alternative 
to (\ref{eq:L24}) is $L_{29}( \Phi^{(5)})\,  =\,  0$  but this 
requires $\,  \Phi^{(5)}$ series to 
$\, 30 \times 1238\,  = \, 37140$ terms.  The change 
of our ODE problem from homogeneous to inhomogeneous
form is a substantial reduction in complexity.}.

The operator $\, L$ defined by $\, L(\phi) = \, 0$
for a given $\, \phi$ is not unique if one does not require $L$ to
be of minimum order.  The advantage of seeking a non-minimal 
order $\, L$ is that the number of unknown
coefficients to be found can drop dramatically.  For 
example, only about 6200 terms are needed 
to obtain the non-minimal $\, L_{29}$ of order 51 and
degree 118 used in~\cite{High}.  The analogous effect occurs 
for the inhomogeneous equation (\ref{eq:L24})
and we find that $\, \Phi^{(5)}$  
series of length only about 5300 terms
are needed when $\, L_{24}$ is chosen of order 42 and
degree 103 leading to polynomials of degree 155 
in $\, E^{(5)}$ in (\ref{eq:E5}) with now 
$\, \kappa \, =\,26$.  Note however, as observed 
in~\cite{experim}, that the integer coefficients in 
non-minimal order operators can be outrageously large
and we expect that Chinese remainder 
reconstruction of the non-minimal order $\, L_{24}$ would
be nearly hopeless.  Instead, the utility of 
a modulo prime non-minimal $\, L_{24}$ lies in its use 
as a recursion device to extend directly generated
(short) $ \Phi^{(5)}$ series to series of sufficient length,
i.e. 26800 terms, so that the minimum order $\, L_{24}$ and 
associated $\, E^{(5)}$ can be found.  Such extension 
requires completely negligible computer resources.

A useful variant of the above approach is to use the 
non-minimal $\, L_{24}$ found as described above
to generate a long series for a solution $\,S_{24}$ 
satisfying $\, L_{24}(S_{24})\, =\,\, 0$.  This series need only 
be about 22300 terms, long enough to enable 
reconstruction of the minimum order $\, L_{24}$.  Finding
the coefficients in $E^{(5)}$ from $\, L_{24}(\Phi^{(5)})$
is then a separate and simpler problem.

A further reduction in the length of the $\, \Phi^{(5)}$ 
series to be directly generated can be obtained 
if one knows a factorization of $\, L_{24}$ with the right
division operator in exact arithmetic.  This 
situation can be realized given the four modulo 
prime series reported in~\cite{High}. We have 
$\, L_{24}\, = \, L_{12}^{(\rm left)}\cdot L_{12}^{(\rm right)}$. 
Knowing $\, L_{12}^{(\rm right)}$ in exact arithmetic allows one to 
obtain any non-minimal order representation of $\, L_{12}^{(\rm right)}$ 
modulo any prime\footnote[9]{This is most easily done by using the
minimum order $\, L_{12}^{(\rm right)}$ to generate 
a series solution $\,S_{12}$ which 
satisfies $\, L_{12}^{(\rm right)}(S_{12}) = \, 0$ but does
not satisfy $\, L_n(S_{12})\, = \,\,0$ for any $ n<12$.  The non-minimal
order $\, L_{12}^{(\rm right)}$ is then found from the series
using, for example, the matrix code described in section~3 of~\cite{experim}.} 
and then $\Psi\, =$  $\, L_{12}^{(\rm right)}(\Phi^{(5)})$
modulo a prime from the directly generated $\, \Phi^{(5)}$.
If we choose our representation of the known $\, L_{12}^{(\rm right)}$ 
as order 18 and degree 42, and the unknown $\, L_{12}^{(\rm left)}$ as 
order 32 and degree 89, then the polynomials in
$E^{(5)}\,  = \,\, L_{12}^{(\rm left)}(\Psi)$ are of degree 199.  This
gives $\,33 \times 90\, +\, 5 \times 200\,  = \, 3970$
unknown coefficients in $\,E^{(5)}$ and
$\, L_{12}^{(\rm left)}$ to be determined implying that
4000 terms of the directly generated $\,\chi^{(5)}$  series
is more than adequate.  Since the series generation of $\,\chi^{(5)}$
described in~\cite{experim} is an 
$\, O(N^4\cdot \ln(N))\, $ process, this
represents a 3-fold reduction in computer time
from that needed using an unfactored $\, L_{24}$ or a 6-fold 
reduction relative to the unfactored $\, L_{29}$ approach.

We have generated the $\tilde{\chi}^{(5)}$ series
to a minimum $\,\,O(w^{4000})\,$ 
for 90 primes $\,p\,<\, 2^{15}$ and from these
generated $\, L_{24}$ modulo a prime as described above.  To 
obtain the exact $\, L_{24}$ is then a problem
of rational reconstruction but it is relatively easy 
from just a few terms to guess a 
normalization factor\footnote[5]{This becomes the value
of the head polynomial at $w\, =0$ and
is $2^{12}\cdot 3^{13} \cdot 5^7\cdot 7^6\cdot 11^4\cdot 23\cdot 29\cdot 7225564279$
= 4235287273136998077435560752320000000.} 
that converts the problem to integer 
reconstruction by the Chinese remainder theorem.  The 
integer coefficients in $\, L_{24}$ are observed 
to typically have very large powers of 2 as factors 
which one can determine by a process
of trial division by $\, 2^k$.  If $\, k$ is chosen too small,
Chinese remainder reconstruction 
with a fixed number of primes might fail
because the unknown coefficient is too large 
while if $k$ is chosen too large there is failure because
the coefficient is no longer an integer.
With 84 primes we find an intermediate $\, k$
range that yields a consistent integer
reconstruction for every coefficient in $\, L_{24}$.  With
90 primes we have a large number
of consistency checks that leave no doubt
that our reconstruction is exact.  We have 
also confirmed the apparent singularity
constraint equations (A.8) in~\cite{experim} 
are satisfied by our reconstructed $\, L_{24}$ in all cases, that is,
19849 satisfied conditions on 22202
non-vanishing coefficients in $L_{24}$.

In view of the ``massive'' calculations required to find $L_{24}$ in exact 
arithmetic it is natural to ask for some further mathematical and numerical 
checks of the  correctness of the operators $\, L_{24}$ and  $\, \, L_{12}^{(\rm left)}$.
First of all we have checked directly that $\, \, L_{12}^{(\rm right)}$
does indeed right divide $\, L_{24}$ in exact arithmetic. 
Secondly, we have confirmed that the exponents of  the operators 
$\, L_{24}$ and $\, \, L_{12}^{(\rm left)}$ are   {\em rational numbers}
and in agreement with our previous massive numerical calculations~\cite{experim}. 
Thirdly, the  minimal order operator $\, L_{29}$ corresponds to an integral of an algebraic 
integrand, and it is therefore, as mathematicians say a {\em Period} 
(``Derived From Geometry''~\cite{bo-bo-ha-ma-we-ze-09}):
$\, L_{29}$ is thus, necessarily, {\em globally nilpotent}. This is a stronger constraint
than being a Fuchsian operator (with integer coefficients) having only rational exponents. 
Since $\, L_{24}$  is a factor of $\, L_{29}$ it must also be globally nilpotent, 
and likewise the  left factor $\, \, L_{12}^{(\rm left)}$ must be globally nilpotent.
We have verified that $\, L_{24}$, $\, \, L_{12}^{(\rm left)}$
and $\, \, L_{12}^{(\rm right)}$\footnote[2]{However, this operator is
automatically globally nilpotent courtesy of its direct  sum construction.} 
are consistent with globally nilpotent operators. This is a very strong indication
that our exact expressions for $\, L_{24}$ and  $\, \, L_{12}^{(\rm left)}$
are  correct.   To check global nilpotence numerically requires one to 
calculate the  $p$-curvature and check that it is zero for almost all primes.
In practice one can obviously only do this for the first few primes.
The primes used for $L_{12}$ were those smaller than 30
while primes less than 10 were used for $L_{24}$.

\subsection{Computational details}

As shown in~\cite{experim} the calculation of a series for
$\tilde{\chi}^{(5)}$ is a problem with computational
complexity $O(N^4\ln N)$. 
In~\cite{experim} we initially calculated $\tilde{\chi}^{(5)}$ to 10000 terms 
which required some 17000 CPU hours on an SGI Altrix cluster with 1.6 GHz 
Itanium2 processors.  From  this single series we could already exactly identify 
a simple right divisor of $L_{29},$ and using this factor we were able
to find a solution modulo a second prime using a series of `just' 5600 terms.
Expanding the series to order 5600 took around 1560 CPU hours on the Altrix cluster.
The series modulo these two primes then sufficed to find a larger right divisor
of $L_{29}$ in exact arithmetic, and using this factor we found that only 4800 terms 
would be required to find solutions for any subsequent primes.
 
Shortly after these developments  a new system was installed by
the National Computational Infrastructrure (NCI) whose National Facility 
provides the national peak computing facility for Australian researchers. This new system is an
SGI XE cluster using quad-core 3.0GHz Intel Harpertown cpus. Our code
runs almost  twice as fast on this facility compared to the Altrix cluster.
We then used this system to calculate the series for $\tilde{\chi}^{(5)}$ to order
4800 (this took about 450 CPU hours) for a third prime, which again allowed us to find 
an even larger right divisor of $L_{29}$ in exact arithmetic.  This larger operator 
reduced the required  number of terms to 4600
and we then calculated a series for a fourth prime to this order using
some 380 CPU hours.  These calculations gave us results for 4 different primes  
and allowed us to  reconstruct  the factor $L_{12}^{(\rm right)}$ in exact arithmetic 
as begun in~\cite{High} and completed here in \ref{exactL1tilde}.

It was only after this that we realised that the inhomogenous equation (\ref{eq:L24})
could be used, as detailed above, to obtain simultaneous solutions for
$L_{24}$ and $E^{(5)}$ using as few as 4000 terms. We then calculated series  for 
$\tilde{\chi}^{(5)}$ to order 4000 for a further 86 primes with each
prime requiring about 215 CPU hours.

\begin{figure}
\begin{center}
\includegraphics[width=12cm]{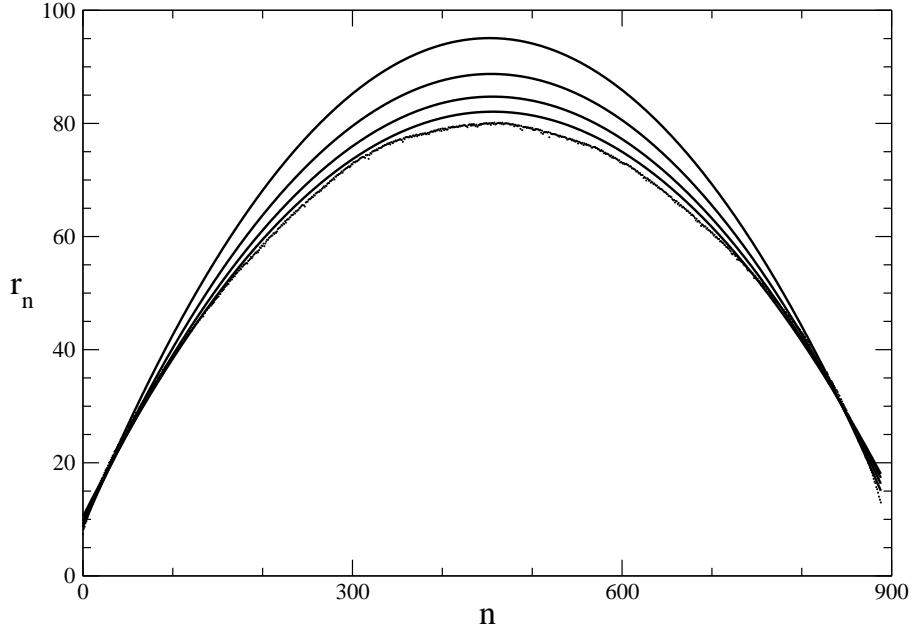}
\end{center}
\caption{\label{fig:L24} Estimates $r_n=\ln (c_n)/\ln (30000)$ for the number of primes 
required to reconstruct the coefficients $c_n$.}
\end{figure}

The above timings make it clear that the reconstruction of $L_{24}$ and $\, E^{(5)}$ 
is a computationally expensive project. The main computational effort is the
direct calculation of the series $\tilde{\chi}^{(5)}$ modulo the required number
of primes. It is therefore of some practical interest to estimate the number
of primes required for the exact reconstruction based only on some partial
reconstruction. We focus here on the coefficients of the head-polynomial of $L_{24}$
and denote by $c_n$ the $n$'th coefficient ($n=0,\ldots,\, 888$) 
after stripping it of any factors of 2 as mentioned above. Then  $r_n=\ln (c_n)/\ln (30000)$ 
is a rough measure of the number of primes needed to reconstruct $c_n$. From our previous 
reconstruction of $L_{12}^{(\rm right)}$ we noticed that the corresponding
$r_n$ are given roughly by a quadratic function of $n$. Thus one can
estimate the number of primes needed to reconstruct the full $L_{24}$ 
from a partial reconstruction since $r_n$ for $n$ near 0 or 888 can
be obtained using many fewer primes.  In Fig.~\ref{fig:L24} the lower `curve' is
the actual data from the coefficients of the head-polynomial of $L_{24}$,
while the upper solid curves are quadratic fits to the
data based on the part data set, from top to bottom, 
$r_n \leq 30,\, 40,\, 50,$ and $60$. The pertinent point being that after 
doing the calculation for some 30 primes we estimated that the
reconstruction was likely to succeed with no more than 100 primes and was
therefore achievable in practice. 

With $L_{24}$ and $\, E^{(5)}$ known in exact arithmetic it is easy to use
(\ref{eq:L24}) to calculate the exact series for $\Phi^{(5)}$ to high order. Specifically our
reconstruction means that we know the coefficients $a_{i,j}$ of the polynomials
in the operator $L_{24}$ and the coefficients of the polynomials $P_{i,j}$ of (\ref{eq:E5})
exactly. The latter allows us to easily calculate the coefficients $e_n$ of $\, E^{(5)}$
by using the (simple) recursive formulae for the elliptic integrals $E$ and $K$. 
From (\ref{eq:L24})  we have explicitly, by equating  the coefficients of $x^n$, that 
(recall that $L_{24}$ is expressed in terms of the differential operator 
$x {\rmd \over \rmd x}$) 
\begin{equation} \label{eq:cn}
\sum_{i=0}^M \sum_{j=0}^D a_{i,j}\, (n-j)^i \, c_{n-j} = Ac_n +B = e_n,
\end{equation}
where $A$ and $B$  are integers (depending on $n$). 
The coefficients $c_n$ of $\Phi^{(5)}$ can thus be calculated recursively
and from (\ref{eq:Phi5}) we can calculate the coefficients of $\tilde{\chi}^{(5)}$
(with the coefficients of $\tilde{\chi}^{(3)}$ calculated using the ODE from \cite{ze-bo-ha-ma-05c}).
We have calculated the coefficients of $\tilde{\chi}^{(5)}$ up to order 8000 and 
they can be found in \cite{http}.

Finally we decided to calculate the minimal order operator $L_{29}$ explicitly; 
it is given implicitly by (\ref{eq:L24}).  This can be done in a variety of ways.
The  most obvious way is to use (\ref{eq:cn}) to extend the series for
$\Phi^{(5)} $ to high enough order ($30\times 1238 = 37140$,  since
the minimal order $L_{29}$ has degree 1237) and then use the matrix 
code of \cite{experim} to calculate the ODE corresponding to $L_{29}$ modulo 
a sufficient number of primes to reconstruct $L_{29}$. However, computationally 
it is easier to first calculate modulo a prime the minimal operator $L_5$ annihilating 
$E^{(5)}$ and then form the product $L_5\cdot L_{24}$ modulo a prime. The minimal
order operator for $L_5$ has degree $4489$ so $6\times 4490 = 26490$ terms
of $E^{(5)}$ is required. Obviously, since the minimal order $L_{24}$ has 
degree 888, the product  $L_5\cdot L_{24}$ has degree 5377 meaning that there is
a common factor of degree 4140, which we must discard in order to calculate
the degree 1237 polynomials of $L_{29}$. The minimal order $L_5$ was
calculated (for each prime) using the matrix code of \cite{experim}. 
The product $L_5\cdot L_{24}$ was then calculated  modulo a prime
using Maple and the common factor can then be divided out modulo a prime.
The bottle-neck in this calculation is the use of the matrix code of \cite{experim}
which has computational complexity $O(N^3)$. It was for this reason that we
chose the ``indirect'' route of going through $L_5$ to get $L_{29}$.

\vskip .5cm

\section{Proof that $\, L_{12}^{(\rm left)}$ does not factorize}
\label{proof}
\vskip .1cm

The factorization of $L_{29}$ defined by $\, L_{29}(\Phi^{(5)}) = \, 0$
described in~\cite{High} relied mostly on the 
testing of series $\, S(w)$ generated by $L_{29}(S) = \, 0$
modulo a prime around $\, w=\, 0$.  If a particular
series is annihilated by an $L_n$ for order $\, n <\, 29$
then $L_n$ right divides $L_{29}$.  Depending 
on the singularity exponents, a series 
solution\footnote[1]{If the solution contains powers
of $\, \ln(w)$ the $\, S(w)$ here is to be interpreted
as the coefficient of the highest power of $\, \ln(w)$.} $S(w) = \,$
$ w^q \cdot (1\, +\alpha_1 \, w\,+\alpha_2 \, w^2\, + \cdots )$ 
with fixed $\, q$, might be uniquely determined by $L_{29}(S) =\,  0$ 
or might contain one or more arbitrary rational coefficients $\alpha_i$.
In the latter case the series may be a generator for a right division 
operator only for a particular choice of constants  $\alpha_i$ and 
the problem then is how these particular values might be found.  
If there is only one arbitrary coefficient $\alpha$ in the series $\, S(w)$ 
an exhaustive search is possible in a modulo prime, $\, p$,
calculation because one need then only test $\, p$ separate series 
with $\, \alpha$ an integer satisfying $ 0 \, \le  \alpha \, <  \, p$.  
If there is more than one arbitrary $\, \alpha_i$ in the series $\, S(w)$ 
such brute force ``guessing'' is  no longer practical due to computational 
time constraints.  It is this that prevented the authors of~\cite{High} from 
deciding whether $\, L_{12}^{(\rm left)}$ is factorizable.  Let us also note 
that we were not able to perform a straight formal calculation factorization
of $\, L_{12}^{(\rm left)}$ using its expression in exact arithmetic and our
attempted calculations failed on a computer with 48 Gb of 
memory\footnote[5]{More precisely the Maple 13 command 
DFactor(*,onestep) was not able to yield a conclusive answer for such 
complicated large order operators. Seeking for a left division of $\, L_{12}^{(\rm left)}$
we had similar inconclusive Maple calculations on $\, {\rm adj}(L_{12}^{(\rm left)})$,
the adjoint operator of $\,L_{12}^{(\rm left)}$.}.

As pointed out in~\cite{High} there is nothing special about the singular point 
$w\, =\, 0$.  And indeed, a series solution about a singular point $w_s \, \ne \, 0$ 
might also lead to a right division operator.  For example, the unique singular series
\begin{equation}
\label{7over4}
\fl
\qquad x^{-7/4} \cdot  (1\, +387\, x/80\, +72103\, x^2/23040\, \ 
  +2054561\, x^3/1597440\, +O(x^4))
\end{equation}
with $\, x=\, w \, -1/4$ is annihilated by an order three operator. 
 With $\, x=\, w\, -1/2$, the unique singular series
\begin{equation}
\label{7over2}
\fl
\qquad x^{7/2} \cdot (1\, -41\, x/6\, +26557\, x^2/792\,  -8692015\, x^3/61776\,\, +\,\, O(x^4))
\end{equation}
yields an order 6 right division operator $\, L_{6}$.  By proceeding through a sequence 
of such solutions, one can eliminate much if not all of the $p$-fold searching
described in~\cite{High} to achieve the factorization $\, L_ {29} \, = \, L_5 \cdot  L_{24}\,  =
 \, L_5 \cdot L_{12}^{(\rm left)} \cdot L_{12}^{(\rm right)}$ 
and the additional factorization of $\, L_{12}^{(\rm right)}$.

The points chosen for series expansion need not be restricted to the rational head
polynomial roots as in the equations (\ref{7over4}) and (\ref{7over2}) above. 
The factor $\, 1\, +3w\, +4\, w^2$, which appeared already in the head polynomial 
of the $\, L_7$ that annihilated $\, {\tilde \chi}^{(3)}$, has ``accidental'' modulo 
prime factorizations for roughly half the primes close to $2^{15}$.  For example, 
with prime $\,p\,=\,\, 32719$, one has $\, 1\,+3w\, +4w^2\,  =\,  4\,  (w-8973)\, (w-31925)$ 
modulo $\, p$.  We can write $ x=\, w\, -w_p$ with $w_p$ either 8973 or 31925 
and obtain solutions about $x =\, 0$ satisfying 
$\, L_{24}(S(x)\cdot \ln^2(x)\, +R(x) \cdot \ln(x)\, +Q(x))\, =\,\,  0$
modulo $\, p$, where $\,R$ and $\, Q$ are regular at $\, x =\, 0$ and the series 
$\, S(x)\, =\, x\,\,  +O(x^2)$ is unique.  Testing shows that $\, S(x)$ is annihilated 
(modulo $\, p$) by an order 3 operator $L_3(x)$.  As we show in general in~\ref{rightdiv}, 
one can obtain from $L_3(x)$ a right division (modulo $p$) operator $L_3(w)$ and then
from multiple modulo prime calculations, an exact right division $L_3(w)$ by the
Chinese remainder theorem.  This reconstructed $L_3(w)$\footnote[2]{It is equivalent 
to the product $\, Z_2 \cdot N_1$ which was shown on general grounds
in~\cite{High} to right divide $L_{29}$.}   has the factor $\, 1\, +3\, w \, +4\, w^2$ in 
its head polynomial in spite of the fact that $w_p$ is clearly not one of the roots 
$(-3 \, \pm \rmi\sqrt{7})/8$  modulo $\, p$, of $\, 1\, +3w+4\, w^2$.

Testing at points other than $\, w=\, 0$ also enables one to exclude certain series 
solutions as generators of right division operators.  For example, the case 3
polynomial\footnote[1]{We follow the notation of Appendix C in~\cite{experim}.}  
$\, 1\, -7\, w\, +5\, w^2\, -4\, w^3$ which is a factor of the head polynomial of 
$\, L_{24}$ has the modulo prime, $\, p_0\, =\,\,  32749$, factorization 
$32745 \cdot (w^2\, +11821\, w\, +10836)(w\, -3635)$.  Define $\,x\, =\, w\, -3635$. 
Then the singular solution modulo $p_0$ about $x =\, 0\,$ is $\, S(x)\, \ln(x)\, +R(x)$ with
\begin{equation}
\label{eq:lnS}
\fl
\qquad S(x)\,  = \, \,\,  x^5\, +13877\, x^6\, +9339\, x^7\,
 +25021\, x^8\, +21884\, x^9\,\, +\,O(x^{10})
\end{equation}
unique and $\, R(x)$ regular at $x =\, 0$.  Testing the series~(\ref{eq:lnS}) 
shows there is no $L_n(S)\,  =\,\,   0$ modulo $\, p_0$ for any $n\, < \,24$.  
Thus there is no operator of order less than 24 that has $ x\, = \, w \, - \, 3635$ 
as a factor (modulo $\, p_0$) of the head polynomial and more generally,
$\, 1\, -7\, w\, +5\, w^2\, -4\, w^3$ as a factor.  This also implies that any solution 
$\, S$ that is singular at some root of  $\, 1\, -7\, w\, +5\, w^2\, -4\, w^3$ and 
satisfying $\, L_{12}^{(\rm left)}(S) =\,  0$, cannot also be a solution of an operator 
of order less than 12 that right divides $\, L_{12}^{(\rm left)}$.  The same conclusion 
is reached for the remaining case 3 and both case 4 polynomials for 
$\tilde{\chi}^{(5)}$ from Appendix C in~\cite{experim}.

If we only knew $\, L_{24}$ or $\, L_{12}^{(\rm left)}$ modulo prime for a few primes, 
the above information about singular solutions associated with case 3 and 4 
polynomials would not be particularly useful for finding or excluding factorization.  
However, with the exact $\, L_{24}$ available one has global information and can 
match series solutions about $ w\,  =\,\,  0$ to solutions about other singular points
of $\, L_{24}$.  In particular, if one can show that every series solution $\, S(w)$ 
about $w=\, 0$ satisfying\footnote[4]{Equivalently every solution of $\, L_{24}$ 
that is not a solution of $\, L_{12}^{(\rm right)}$.} $\, L_{12}^{(\rm left)}(S)\,  =\, \, 0$ 
is singular at some root of the case 3 or 4 polynomials then one has proved that 
$\, L_{12}^{(\rm left)}$ does not factorize. Our demonstration that this is the case 
uses the singular point $w_s= \, 0.15853 \, \cdots $ which is a root of 
$\, 1\, -7\, w\, +5\, w^2\, -4\, w^3$.  This is a particularly convenient point as it is the 
closest root of the head polynomial of $\, L_{24}$ to both $w=\, 0$ and $ w=\, 1/4$.

We begin the demonstration by studying the two linearly independent solutions
of the form $S_i \, =\, A_i(w) \cdot  \ln^3(w)$ plus terms with lower powers of $\, \ln(w)$.  
The two series
\begin{eqnarray}
\label{eq:A1}
A_1\, & = &\,\,  9\, w\, +261\, w^3\, +1845\, w^4\, +7046\, w^5\, +42771\, w^6\,
 +145980\, w^7\, \nonumber \\
&&   +785528\, w^8\, +2536628\, w^9  \,  +12800309\, w^{10}\, +38627228\, w^{11}
\,\nonumber \\
&&   +187738058\, w^{12}\, +\, \cdots \,\,  +\, \alpha_{1,n} \cdot w^n \, + \,  \cdots,
\end{eqnarray}
\begin{eqnarray}
\label{eq:A2}
A_2 \, &=& \, \, 27\, w^2 \, +102\, w^{3}\, +270\, w^{4}\, +2164\, w^{5}\, +5532\, w^{6}\, 
+43722\, w^{7}\, \nonumber \\
&&   +132130\, w^{8}\, +922108\, w^{9}\, +3158590\, w^{10}\, +19690882\, w^{11}\,\nonumber \\
&&   +72977164\, w^{12}\, +\, \cdots \,  \,  +\, \alpha_{2,n} \cdot w^n \, + \,  \cdots,
\end{eqnarray}
satisfy $\, L_{24}(A_i)\,  =\,\,  0$ {\em but not} $\, L_{12}^{(\rm right)}(A_i)\,  =\,\,   0$
and a ratio test shows they have radii of convergence $\, |w| =\, w_s\,=\, 0.15853\cdots$. 
Thus both $A_1$ and $A_2$ are singular at $ w= \, w_s$ and cannot
be generators of an $L_n$, $\, n <\, 24$, that right divides $\, L_{24}$.  Whether a 
linear combination of $A_1$ and $A_2$  leads to a right division
operator is now determined as follows.

Near $ x =\, 0$ where $x \,= \, w_s\, -w$ the $\, A_i$ must be of the form
$A_i\,  = \, \, B_i \cdot f(x)\cdot  \ln(x)\, +g_i(x)$ where $f(x)$
and the $g_i(x)$ are all regular at $\, x\, =\, 0$.  The series 
$\, A(w)\, \propto \,  B_2 \, A_1(w) \, \, -B_1\, A_2(w)\, $ will be 
regular at $ w= \, w_s$ and if the amplitude ratio $B_1/B_2$
is rational, then $A(w)$ might be a
candidate generator for a right division operator. In principle, the amplitudes 
$B_i$ could be found by matching series solutions 
about $x=\, 0$ to those about $ w=\, 0$ but since
we only need the $B_1/B_2$ ratio, a simpler procedure that utilizes only the
coefficients in (\ref{eq:A1},\ref{eq:A2}) is possible.  We note that of the
remaining singularities of $A_i$, the nearest to 
$ w=\, 0$ are at $|w|=\,1/4$. This implies
that $\, B_1/B_2\,  =\, \alpha_{1,n}/\alpha_{2,n}$, a
ratio of coefficients from (\ref{eq:A1},\ref{eq:A2}), to an accuracy of
order $(4\, w_s)^n \approx \, 0.634^n$.  Thus the
problem reduces to searching for a (small) rational
$B_1/B_2$ from a sequence that converges exponentially.  This can be done by 
expressing $\alpha_{1,n}/\alpha_{2,n}$
as a continued fraction.  We observe that, for $\, n$ greater
than some fixed $n_0$, a particular term in the 
continued fraction grows exponentially
which is a clear indication that in the limit $n\, \rightarrow \, \infty$
the continued fraction terminates
and is the (small) rational $B_1/B_2\,  =\, -637/228$.  
Our result for the linear combination series is then
\begin{equation}
\label{Aw}
\fl \qquad A(w)=  2052w +17199 w^2 +124482 w^3 +592650 w^4
   +2984956w^5 +O(w^6)
\end{equation}
which is confirmed to have a radius of convergence $|w|\, =\, 1/4$.  However,
testing the series (\ref{Aw})   shows it is not annihilated by
any $\, L_n$, $\, \,\, n<\, 24$, and thus we have
eliminated the only possible linear combination
candidate series for a right division operator.

While direct testing is an easy way to exclude $\, A(w)$ in (\ref{Aw}), such direct testing
is impractical\footnote[3]{This was the problem encountered in~\cite{High}.} 
as a general method for excluding the many possible series that arise in the remaining 
part of our proof. Instead we supplement direct testing by
a method that relies on the global information provided by the exact $\, L_{24}$.  
As an illustration of this method, consider again $\, A(w)$.  If one matches (\ref{Aw})  
to series about $w =\, 1/4$ one finds that $\, A(w)$ contains, as the
leading logarithmic function, $A_s\cdot \ln ^2(y)$ where $2\, y =\, 1\,-4w\, $ and
\begin{equation}
\label{As}
\fl \quad A_s=   1 -21469\, y/640 -1489293\, y^2/81920 +229328363\, y^3/10485760 + O(y^4)
\end{equation}
A ratio test on the coefficients in (\ref{As}) shows $A_s$
has radius of convergence $|y| =\, 1/2\,-2\, w_s$ and
thus is singular at $ w = \, w_s$.  This in turn implies that 
there are branches of the function $A(w)$
on which it is singular\footnote[2]{These singularities
can be reached, for example, 
by following a path along the real axis from $w=\, 0$ to $\, w$ just
less than $\, 1/4$, circling $w=\, 1/4$ any number $\, N$ of times, 
and then moving back to $w=\, w_s$ along the real axis.  Since  $A_s$  
multiplies the leading (unique) logarithmic singularity at 
$ w=\,1/4$, there cannot be cancellation 
of the $w=\, w_s$ singularity of $A_s$ on all branches 
distinguished by $\, N$.} at $w=\, w_s$. By forming the linear 
combination $\, A(w)\, \propto \,  B_2 \, A_1(w) \, -B_1\, A_2(w)$
we only succeeded in forcing an ``accidental'' cancellation
of the singularity at $w =\, w_s$ on the principal 
branch of the function.  The singularity remains on at least some 
other branches and thus $A$ is excluded as a generator of a 
right division operator for exactly the same reason as $\, A_1$ and $\, A_2$.

The above argument has excluded, as generators of right division operators, 
eight of twelve linearly independent solutions satisfying$\, L_{12}^{(\rm left)}(S) =\, 0$.  
These we take to be $C_1\,  \propto \, L_{12}^{(\rm right)}(A)$ with $\, A$ from (\ref{Aw})  
and $C_2\, \propto \, \, L_{12}^{(\rm right)}(A_2)$ with $A_2$ from (\ref{eq:A2}) plus 
the six series with leading $\, \ln(w)$ dependence $\, C_i \cdot \ln^p(w)$,
 $p=\, 1,\, 2$ and $3$.  Explicitly,
\begin{eqnarray}
\label{C1w}
&& \fl \qquad C_1(w)\, =\,\,  w^6\, -444\, w^7/11\,
+275773109\, w^{8}/129360\, +19252320091\, w^{9}/194040 \nonumber \\
&&  -964738631897\, w^{10}/388080 -2082457681309\, w^{11}/27720\,\,\nonumber \\
&&       +17517580633073581\, w^{12}/17075520\,+\,O(\, w^{13}),
\end{eqnarray}
\begin{eqnarray}
\label{C2w}
&& \fl \qquad C_2(w)\,  = \,\,w^6\, +403206\, w^{7}/1661\, -13446782071\, w^{8}/19533360\,
\nonumber \\
&&   -2413114741889\, w^{9}/29300040\, -4359267083039 \, w^{10}/1065456\,
\nonumber \\
&&  -875856906689449\, w^{11}/4185720\,\nonumber \\
&&   +23619065101886078533\, w^{12}/2578403520
\, +\, O(w^{13}).
\end{eqnarray}
Because $\, L_{12}^{(\rm right)}$ does not have the factor $ w\,-w_s\, $ in its 
head polynomial, the $\, C(w)$ functions carry the same $w\, =\, w_s$ 
singularities as the $\,A(w)$ from which they have been generated.  Thus
even though the series $C_1(w)$ has radius of convergence $|w|=\, 1/4$, 
the analytically continued function $C_1(w)$ is still singular at $w=\, w_s$ 
on some other branches.  The series $C_2(w)$ has radius of convergence 
$|w|=\, w_s$ and is already singular at $w=\, w_s$ on the principal branch.

It was shown in~\cite{High} that the remaining four solutions satisfying
$\, L_{12}^{(\rm left)}(S)\,  =\,\,   0$ are of the form $C_i(w) \cdot \ln(w)\, +D_i(w)$ 
and $C_i(w),\,\,   i=\, 3,\, 4$, with the $\, C_i$ and $\, D_i$ regular at $\, w=\, 0$.  
These two $\,C_i$, together with the two in (\ref{C1w}, \ref{C2w}), are
linearly independent and can all be generated from the coefficient of $\ln(w)$ in
$\, L_{12}^{(\rm right)}(S_{24})$ where $\,S_{24}$ contains four arbitrary
constants and is of the form 
$S_{24}\,  = \,F(w)\cdot \ln^2(w)\,+G(w)\cdot \ln(w)\,+H(w)$ 
with $F,\, G$ and $H$ all regular at $w\, =\, 0$.  We demand that $\, F$ 
satisfies\footnote[3]{For the explicit $ F$ in (\ref{F}), it happens that 
$\, \rmd F/\rmd\beta_1$ is annihilated by an $\, L_{11}$ and 
$\, \rmd F/\rmd\beta_2$ by an $L_9$.  This reduction from 
$\, L_{12}$ plays no role in our subsequent arguments.}
$\, L_{12}^{(\rm right)}(F) =\, 0$.  This guarantees that $\, C(w)\cdot \ln(w)$ 
is the leading logarithm in $\, L_{12}^{(\rm right)}(S_{24})$ and 
simplifies the subsequent analysis.  Only $\, F$ and $\,G $
are relevant for determining $\,C$ and a possible choice is
\begin{eqnarray}
\label{F}
&&F \, =\,\,\,  \beta_1 \cdot  \big(3177 \, w \, -174840\, w^{4}\,
 -817828\, w^{5}\, -5829558\, w^{6}\, \nonumber \\
&& \quad
 -25983762\, w^{7}\, -142882882\, w^{8}\, 
	        -620769318\, w^{9}\, -3086072424\, w^{10}\,
 \nonumber \\
&& \quad -13199839762 \, w^{11}\,
 -62214586728\, w^{12}\,  
	       \,  +\, O(\, w^{13})\big)\,  \nonumber \\
&& \quad 	   +\beta_2 \cdot \big(3177\, w^{3}\, +13803\, w^{4}\, +74932\, w^{5} \,
 +287997 \, w^{6}\, \nonumber \\
&& \quad +1265280\, w^{7}\, +4296418\, w^{8}\, 
	        +17162736\, w^{9}\, +48945231 \, w^{10}\, 
 \nonumber \\
&& \quad +173557768\, w^{11}\, +284486847\, w^{12}\, 
	        +\, O(\, w^{13})\big),
\end{eqnarray}
and
\begin{eqnarray}
\label{G}
&&G\, = \, \, \,
\beta_1 \cdot \big(1604883673\, w^{8}/210\, +2823208099\, w^{9}/105 
\nonumber \\
&& \qquad +47115755881\, w^{10}/140
	        +782148892459\, w^{11}/630+O(\, w^{13})\big)
\nonumber \\
&& \quad -\beta_2 \cdot \big(366106439\, w^{8}/1050+1576821038\, w^{9}/525
 \nonumber \\
&& \qquad +9206778909\, w^{10}/350
	        +1049578781449\, w^{11}/6300+O(\, w^{13})\big)\nonumber \\
&& \quad +\beta_3 \cdot \big(35\, w^{6}\, +1223\, w^{8}\, +1852\, w^{9}\,
 +36064\, w^{10}\, \nonumber \\
&& \qquad +96388\, w^{11}\, +O(\, w^{13})\big)\nonumber \\
&& \quad 
	   +\beta_4 \cdot \big(105\, w^{7}\, +304\, w^{8}\, +3536\, w^{9}\,
 +10192\, w^{10}\,\nonumber \\
&& \qquad  +79089\, w^{11}\,\, +O(\, w^{13})\big),
\end{eqnarray}
where the $\, \beta_i$ are arbitrary constants.  We will now show that 
no choice of these constants can yield a 
$\,C \,= \,  C(\beta_1,\, \beta_2,\, \beta_3,\, \beta_4)$, defined by
\begin{eqnarray}
\label{def}
L_{12}^{(\rm right)}(F \cdot \ln^2(w)\,\,+G \cdot \ln(w)\, +H)\,\,  
=\,\,\, C \cdot \ln(w)\,\,+D,
\end{eqnarray}
that is a generator for a right division operator of $\, L_{12}^{(\rm left)}$.  
The argument is essentially that given above for the exclusion of $C_1$
and $C_2$ in (\ref{C1w}, \ref{C2w}).  In fact the demonstration has already
been partially completed since, in terms of 
$\, C(\beta_1,\, \beta_2,\, \beta_3,\, \beta_4)$, $\, C_1\, \propto \,  C(0,0,3,4)$ 
and $C_2 \, \propto \, C(0,0,453,1744)$.  Since $\, F$ in (\ref{F}) satisfies
$\, L_{12}^{(\rm right)}(F)\, =\,\,  0$ it is not singular at 
$w\, = \, w_s=\, 0.15853 \cdots$, neither is $\, L_{12}^{(\rm right)}(F \cdot \ln^2(w))$.  
Thus it suffices to investigate $\, G $ and if every $G$ is singular at $\, w =\, w_s$ 
then so is $\, C$ defined by (\ref{def}) and we have proved $\, L_{12}^{(\rm left)}$ 
does not factorize.

A ratio test on the series coefficients in (\ref{G}) shows the generic $G$ has 
radius of convergence $|w|=\, w_s$ and thus is singular at $ w=\, w_s$.  
But by the same analysis that led from the series (\ref{eq:A1}, \ref{eq:A2}) 
to the linear combination (\ref{Aw}), we can construct three 
$G = \, G(\beta_1,\, \beta_2,\, \beta_3,\, \beta_4)$ each of whose radius of 
convergence is $\, |w|=\, 1/4$.  These are
\begin{eqnarray}
\label{105}
G(105,0,0,-1182781), \quad G(0,525,0,-1443727),\quad  G(0,0,3,4).
\end{eqnarray}
To the remaining linearly independent $G =\, G(0,0,0,1)$ one can add any 
combination of the three in (\ref{105}) but this will not change its radius
of convergence from $|w|=w_s$ and remove the singularity at $w\, =\, w_s$.  
In this sense $ \, G(0,0,0,1)$ is equivalent to $\, G(0,0,453,1744)$
which corresponds to $C_2$ via (\ref{def}) and is excluded as
a generator of any right division operator.

To determine the behaviour of the three $G$ functions in (\ref{105}) in the 
vicinity of $y =\, 0$ where $2y\, =\, 1\, -4w$ we match the series\footnote[1]{It follows 
from $\, L_{24}(F \cdot \ln^2(w)\, +G \cdot \ln(w)\, +H)\, =\, 0$ by analytic 
continuation around the $w=\, 0$ singularity that also
$\, L_{24}(2 \, F \cdot \ln(w)\, +G)= \, 0$.} $ \, 2\, F \cdot \ln(w)\, +\, G$
in $w$ about $w =\, 0$ to solutions $\, S$ satisfying $\, L_{24}(S) =\, 0$ 
about $y=0$.  Since $\, L_{12}^{(\rm right)}(F)=\, 0$ one can show that 
$\, F$ (and $F\cdot \ln(w)$) can contain only the first power of $\, \ln(y)$ 
near $y=\, 0$. Any $\, \ln^2(y)$ or $\, \ln^3(y)$  we find in the matching 
$\,S$ can only come from $\, G$ in the combination solution 
$\, 2 \, F\cdot \ln(w)\, +G$.  Our matching shows the leading logarithmic 
dependencies of the three $G's$ in (\ref{105}) are respectively
\begin{eqnarray}
\label{3420025}
&&(-3420025/8192/\pi^2) \, A_s\, \ln^3(y), \quad \quad
 (-2473625/16384/\pi^2) \, A_s\, \ln^3(y), \nonumber \\
&& \qquad (-125/16384/\pi^2) \, A_s\, \ln^2(y)	
\end{eqnarray}
where $A_s$ is given by (\ref{As}).  A linear combination of the first two $G's$ 
in (\ref{105}) can be constructed to eliminate the leading $A_s \, \ln^3(y)$ 
shown in (\ref{3420025}) and we find the resultant 
$G(494725,-6840050,0,13236968029)$ has $\, A_s\cdot \ln^2(y)$ as the 
leading logarithmic singularity.  Clearly this remaining singularity can now
be eliminated by forming a linear combination with the last G 
in\footnote[5]{The surprise, at least for us, is that all of the necessary 
combinations can be formed with rational amplitudes.} (\ref{105}).
In summary, we have generated the three 
$\,  G(\beta_1,\, \beta_2,\, \beta_3,\, \beta_4)$ combinations
\begin{eqnarray}
\label{494725}
&&G(494725,-6840050,15276842775,33606091729),\quad 
 \nonumber \\
&&G(0,525,0,-1443727),\quad \quad \quad  G(0,0,3,4)
\end{eqnarray}
which are irreducible in the sense that no further superposition can 
eliminate the  $\, A_s\cdot \ln^3(y)$ and $\, A_s\cdot \ln^2(y)$
singularities of the last two $G's $ while the first $G$ is unique in that 
it contains neither $\, A_s\cdot \ln^3(y)$ nor $\, A_s\cdot \ln^2(y)$ .  
The last $G$ in (\ref{494725}) we have already identified as being
associated with $C_1$.  We associate the middle $\, G$ in (\ref{494725})  
with $C_3\propto \,  C(0,525,0,-1443727)$ which cannot be a generator 
of a right division operator of $\, L_{12}^{(\rm left)}$ for exactly the same 
reason as $\, C_1$.  The first $\, G$ in (\ref{494725}) we associate with 
$C_4\, \propto \,  C(494725,-6840050,15276842775,33606091729)$
and test $\, C_4$ directly. We find there is no operator satisfying 
$\, L_n(C_4) = \, 0$ with $n <\, 12$ and this completes our proof that 
$\, L_{12}^{(\rm left)}$ does not factorize.  The explicit new 
$C_i$, supplementing those in (\ref{C1w},\ref{C2w}) are
\begin{eqnarray}
&&C_3(w)\,  = \,\,w^6+281575923\, w^7/34167793\, 
\nonumber \\
&&\quad  +48755202697119\, w^8/8371109285
\nonumber \\
&&\quad \,  +788146152364265\, w^9/5022665571 \nonumber \\
&&\quad  \,  -48321460210711729\, w^{10}/33484437140
\nonumber \\
&&\quad \, -1046678480403963299\, w^{11}/4783491020 
\nonumber \\
&&\quad \, -4705373665277858926411\, w^{12}/3683288085400\,
 +O(\, w^{13}),
\end{eqnarray}
\begin{eqnarray}
&&C_4(w)\, =\,\,w^6\,-124536\, w^{7}/649\,+2840488261\, w^{8}/508816\,
 \nonumber \\
&&\quad +39013193251\, w^{9}/254408\,
	        -24532098411899\, w^{10}/9540300\,
\nonumber \\
&&\quad  -6082145734733\, w^{11}/68145\,
	        \nonumber \\
&&\quad -179169570633725593\, w^{12}/314829900\, +\,O(\, w^{13}).
\end{eqnarray}

\subsection{More on the structure of the differential operator $L_{12}^{\rm (left)}$ }

Once the differential operator $L_{12}^{\rm (left)}$ has been proved to be
irreducible, one may wonder whether this high order differential operator
can nevertheless be built from factors of lower order.
A high order differential operator can be irreducible and still  
result from ``operations" involving differential operators of lower order since
it may be a symmetric power of a lower order differential operator
or a symmetric product of two (or more)  lower order differential operators.

The symmetric $n$'th  power of a differential operator $L_q$ is
the differential operator whose corresponding ODE annihilates a
generic linear combination of the $q$ solutions of $L_q$ to the power $n$.
The symmetric  $n$'th  power of a differential operator $L_q$ of order $q$
has order ${ (q+n-1)! \over (q-1)!n!}$.

The symmetric product of the differential operators $L_{q_1}$ and $L_{q_2}$
of orders $q_1$ and $q_2$, respectively,  is the differential operator whose 
ODE annihilates the product of a generic linear combination of the $q_1$ solutions
of $L_{q_1}$ and a generic linear combination of the $q_2$ solutions of  $L_{q_2}$.
This symmetric product is of order\footnote[2]{The order $q=q_1 \cdot q_2$ is
for the generic case. In general the order of the symmetric product
is $ q_1 + q_2-1 \le q \le q_1 \cdot q_2$.} $q_1 \cdot q_2$.

We use the notation $[w^p]$ to indicate a  series that starts as 
$w^p \,(const. + \cdots)$. In \cite{High} it was shown that the formal 
solutions of $L_{12}^{\rm (left)}$ at $w=0$ follow this scheme:
There are two sets of four solutions ($k=6,\, 7$)
\begin{eqnarray}
\label{series7}
&& [w^k] \, \ln(w)^3\, + [w^5] \, \ln(w)^2 \,+
[w] \, \ln(w) \, + [w],  \,  \nonumber \\
&& [w^k] \, \ln(w)^2\, + [w^5] \, \ln(w)\, + [w],   \nonumber \\
&& [w^k] \, \ln(w) \, + [w], 
\qquad \quad \hbox{and} \quad  \qquad  [w^k] 
\end{eqnarray}
and two sets of two solutions ($k=8,\, 9$)
\begin{eqnarray}
\label{series9}
 [w^k] \, \ln(w) \, + [w], \qquad\quad 
\hbox{and} \qquad \quad   [w^k], 
\end{eqnarray}
\begin{eqnarray}
\label{series8}
 [w^k] \, \ln(w) \, + [w], 
\qquad \quad  \hbox{and}\quad  \qquad  [w^k]  
\end{eqnarray}
We denote by $BLn$ a set of solutions such as (\ref{series7}) containing
$n+1$ solutions with a logarithmic solution of maximal degree $n$.
For the scheme above, we thus have two $BL3$ blocks and two $BL1$ blocks,
and in each block there is  also a non-logarithmic solution.

We first consider the possibility that $L_{12}^{\rm (left)}$ is a symmetric power of
an operator of lower order. It is straightforward to see that the only possibility
is that $L_{12}^{\rm (left)}$ could be a symmetric eleventh power of a differential
operator of order two with one $BL1$ block.
This possibility is ruled out, since there is no $BL11$ block in
the solutions of $L_{12}^{\rm (left)}$.

Next for the possibility that $L_{12}^{\rm (left)}$ is a symmetric product of 
differential operators  of lower order (we consider only the cases where
the product has the maximal order). There
are three cases to consider:  The symmetric product of differential operators of
orders two and six (configuration denoted $2 \cdot 6$), three and four
(configuration $3 \cdot 4$), or two, two and three (configuration $2 \cdot 2 \cdot 3$).

The symmetric product of two differential operators $L_1$ and $L_2$
containing the  blocks $BLn_1$ and $BLn_2$, respectively,  should contain in its
solutions the block $BLn$ with $n=n_1+n_2$. Since the differential operator 
$L_{12}^{\rm (left)}$ contains two $BL3$ blocks  the case $2 \cdot 2 \cdot 3$ is ruled out.

Let us detail the compatibility of the case $3 \cdot 4$ at $w=0$.
With two $BL3$ blocks in $L_{12}^{\rm (left)}$ the only possibility is that
the order three differential must have one $BL2$ block:
\begin{eqnarray}
\label{solBL2}
&& S_1 \, \ln(w)^2 \, + S_{11} \, \ln(w) \, + S_{10},  \nonumber \\
&& S_1 \, \ln(w) \, +  S_{20},  \nonumber \\
&& S_1 
\end{eqnarray}
and the order four differential operator must have
two $BL1$ blocks:
\begin{eqnarray}
&& T_1 \, \ln(w) \, + T_{10},  \nonumber \\
&& T_1,   \\
&& V_1 \, \ln(w) \, + V_{10},  \nonumber \\
&& V_1  
\end{eqnarray}
It is a simple calculation to form the product of a combination from the
set $BL2$ with a combination of the solutions from the two sets $BL1$.
One obtains:
\begin{eqnarray}
\fl \quad S_1 \cdot T_1 \, \ln(w)^3 + \left( S_1 \cdot T_{10} + S_{11} \cdot T_1 \right)\, \ln(w)^2
+ \left(S_{10} \cdot T_1 +S_{11} \cdot T_{10} \right) \ln(w)+ S_{10} \cdot T_{10}, \nonumber \\
\label{solST2}
\fl \quad S_1 \cdot T_1\, \ln(w)^2 \, + S_{11} \cdot T_1 \, \ln(w)+ S_{10} \cdot T_1, \\
\label{solST1}
\fl \quad S_1 \cdot T_1\, \ln(w) \, + S_{20} \cdot T_1, \\
\fl \quad S_1 \cdot T_1
\end{eqnarray}
 and a similar set of four solutions with $V$'s instead of $T$'s. 
These eight solutions correspond to the two $BL3$ occurring for $L_{12}^{\rm (left)}$
at $w=0$. One obtains also
\begin{eqnarray}
\label{solST2p}
&& S_1 \cdot T_1\, \ln(w)^2 \,+ \left( S_1 \cdot T_{10}+T_1 \cdot S_{20} \right) \ln(w)
+  S_{20} \cdot T_{10},  \\
\label{solST1p}
&& S_1 \cdot T_1 \, \ln(w) \,+ S_1 \cdot T_{10}
\end{eqnarray}
and two other solutions where $V$'s replace $T$'s.
Subtracting (\ref{solST2p}) from (\ref{solST2}) and
(\ref{solST1p}) from (\ref{solST1}), one obtains
\begin{eqnarray}
\label{solLast}
&& \left( S_1 \cdot T_{10}+T_1 \cdot S_{20} -T_1 \cdot S_{11} \right)\, \ln(w) \,+
\left( S_{20} \cdot T_{10}-S_{10} \cdot T_1 \right), \\
&& S_1 \cdot T_{10} - T_1 \cdot S_{20}.
\end{eqnarray}
Note that in a set of solutions such as $BL2$ above, the series
$S_{11}$ depends on $S_{20}$ and $S_1$ and can be expressed as
\begin{eqnarray}
S_{11} \,=\, \alpha \, S_1 + 2\, S_{20}
\end{eqnarray}
The coefficient 2 is generic for any order three ODE and $\alpha$ is a constant
that depends on the ODE at hand. Inserting this $S_{11}$ in (\ref{solLast}),  
we can arrange to have the non-logarithmic series be the same as the
series in front of the log. This is then one of the $BL1$ blocks occurring in
$L_{12}^{\rm (left)}$. The second   $BL1$  block is obtained by considering
(\ref{solLast}) with $V$'s instead of $T$'s.

We have thus shown that under the hypothesis that $L_{12}^{\rm (left)}$ is a symmetric
product of two factors, the scheme of solutions at $w=0$ is compatible with
the configuration $3 \cdot 4$. Similar calculations show that the configuration
$2 \cdot 6$ is also compatible and in this case the order two operator has a $BL1$ block
and the order six operator has two $BL2$ blocks.

Next we must check for each configuration ($3 \cdot 4$ and $2 \cdot 6$)
whether or not the symmetric product is compatible with the scheme of solutions for
$L_{12}^{\rm (left)}$ at those other singularities  containing enough logarithmic 
solutions. We recall that the scheme at $w=\infty$ is the same as the scheme at $w=0$.
The scheme of solutions of $L_{12}^{\rm (left)}$ at the singularity $w=1/4$ is
one $BL3$ and four $BL1$ blocks. One sees immediately that the configuration $3 \cdot 4$
is ruled out. For any set of solutions that we attach to the order three and order
four differential operators, we end up with either more than one $BL3$ or at
least one $BL2$. The  configuration $2 \cdot 6$ is acceptable, since in this case 
there can be one $BL1$ block in the order two operator and one $BL2$ 
plus three $BL0$ blocks in the order six operator.
It remains to be seen whether or not the configuration $2 \cdot 6$ is compatible with the
scheme of solutions at the point $w=-1/4$, which is two $BL2$, one $BL1$ and four
$BL0$ blocks. Our checks show that the configuration $2 \cdot 6$  is ruled out.

In conclusion we have shown that the differential operator $L_{12}^{\rm (left)}$
isn't a symmetric $n'$th power of a lower order operator nor is it
a symmetric product of two (or more) operators of orders $q_1$ and $q_2$ under
the hypothesis that the order of the symmetric product reaches its maximum
value $q_1 \cdot q_2 =12$. It should be noted that our considerations regarding
symmetric powers/products (unlike the question about the factorization of
$L_{12}^{\rm (left)}$) do not require knowledge about $L_{12}^{\rm (left)}$ in
exact arithmetic. The block structure of the solutions can be obtained from
the operator modulo a prime.

\section{Analytic continuation of $\, \Phi^{(5)}$  and its behaviour at $\, w \, = \, 1/2$}
\label{analyt}
\vskip .1cm 

Our analytic continuation of $\, \Phi^{(5)}$  is limited to paths that follow the
real $\, w$ axis on the intervals $[0,\, 1/4]$, then $[1/4,\, (3-\sqrt {5})/2]$, 
and finally $[(3-\sqrt {5})/2),1/2]$.  We allow any number of half-integer turns 
about the ferromagnetic point $\, w=\, 1/4$, that is, rotation by any angle 
$\, \theta = n\, \pi$ with $\, n$ odd.  This is followed by any rotation 
$\, \theta = m\, \pi$, $\, m$ odd, around $\, w\, =(3-\sqrt {5})/2$.  This point 
is one of the $s$-plane circle singularities discussed by Nickel~\cite{nickel-99,nickel-00}.  
The point $\, w =\, 1/2$ also maps onto the $|s|=\, 1$ circle but it is not a 
singularity of $\tilde{\chi}^{(5)}$  when it is approached on any path that 
does not leave the principal disk $|s| \le \, 1$.  In terms of our $n,m$ paths 
in the $w$-plane, the combinations $n=\, m=\, \pm 1$  are such principal disk
constrained paths.  Every other $n,m$ combination is a path that reaches 
$\, w=\, 1/2$ on another branch.

Our starting point for the analytic continuation is the $\, \Phi^{(5)}$  series 
expansion about  $\, w=\, 1/4$ in \ref{ferro}.  For $\, 0\, \le\,  w <\, 1/4$, $2y=\, 1\,-4w$  
is positive real and the half-integer turns about $\, w=\, 1/4$ that bring one to
$ \, w\, >1/4 $ simply requires the replacements
\begin{eqnarray}
\label{yy}
y \, \longrightarrow \,  \,  -y,  \qquad \quad   \ln(y/4) \, \longrightarrow 
\,  \,  \,  \ln(|y|/4)\,  -\rmi  n  \pi, 
\end{eqnarray}
in (\ref{A3}) with $n$ odd to be understood.  The new (\ref{A3}) series generated 
with the replacements (\ref{yy}) is now to be matched to series in $\, z$ where
$\,z=\, 3\, -\sqrt{5} -2\, w =\,  5/2\,  -\sqrt{5} \, -y$.  Although direct matching is 
possible, considerable improvement in the utility of the $\, y$ series results
by first making an Euler transformation by the replacement $y \rightarrow \,  y/(1-y)$.  
This has the effect of bringing the $w \, =\, (3\, -\sqrt{5})/2\, \approx \, 0.382$
singular point closer to $\, w=\, 1/4$ while moving $w_s\, \approx \, 0.1585$
further away.

To generate $\, \Phi^{(5)}$  series in $\, z$ we must first analytically continue
the elliptic integrals in (\ref{eq:E5}).  The replacements required for $\,w >1/4$, 
with the same $\, n$ as in (\ref{yy}), are
\begin{eqnarray}
&&K(4w) \, \longrightarrow \, \, \,    u[K(u)\, +\rmi \, n K(u')],
 \nonumber \\
\label{eq:Enew}
&&E(4w)\, -K(4w) \, \longrightarrow \, \,  \,  [E(u)\, -K(u)\, -\rmi \,n  E(u')]/u
\end{eqnarray}
where $u=\,\,  1/(4w)$ and $\, u'\,  =\, \, \sqrt{1-u^2}$.  The new elliptic 
integrals (\ref{eq:Enew}) are easily developed as series in $\, z$
from their defining differential equations set up as recursion relations.  The remaining 
step of finding a particular integral of (\ref{eq:L24}) together with all homogeneous 
series solutions is also straightforward.  The ODE numerical recursion in $z$ is not 
as unstable as the recursion in $\, y$ noted in \ref{ferro}.  Here one loses only about 
a factor 10 in relative accuracy for each two orders in $\,z$.

The matched series in $z$ is of the form $\, A(z)\cdot \ln(z)\, +B(z)$ where $A$ and $B$ 
are regular at $z=\, 0$.  Since $\tilde{\chi}^{(1)}$ and $\tilde{\chi}^{(3)}$ are not singular 
at this point, we can identify the singularity in $ \,\Phi^{(5)} $ with that in $\tilde{\chi}^{(5)}$.  
We find the leading singular term is\footnote[2]{This is based on floating point 
results to about 300 digit accuracy.}
\begin{equation}
\label{26}
\fl \qquad [(1+10 n^2+5n^4)/16] \cdot [(25/693)(5-\sqrt{5})(2+\sqrt{5})^{11}/(2^{22}\, \pi^2)]
\cdot  z^{11} \, \ln(z).  
\end{equation}
The $n=\, 1$ singularity, when mapped to $s$-plane variables, agrees with the sum 
contribution  $\tilde{\chi}^{(5)}_{0,1}\, +\tilde{\chi}^{(5)}_{0,-1}$ from 
equation~(14) in~\cite{nickel-99}.  The new result in (\ref{26}) is the branch 
dependent multiplicity $(1\,+10\, n^2\,+5\,n^4)/16$.

To most clearly identify a possible $\, w\, =\, 1/2$ singularity in the $\, A(z)$ and $\, B(z)$ 
series we make another Euler transformation with the replacement 
$\, z \rightarrow \,  z/(1\, +6\,z/5)$.  This moves the known singularity at $\, w\, =\, 1/4$ 
to the new $\, z=\, 5\,(\sqrt{5}-1)/16\, \approx \, 0.386$ and the potential 
singularity at  $\, w\, =\, 1/2$ to the new $\, z=\, -5\, (16\,-5\, \sqrt{5})/131\, \approx \,-0.184$. 
There is another potential singularity at a complex root of the case 4 polynomial
$\,1\,-w\,-3w^2\, +4w^3$. This maps to the new $\, |z|\, \approx \, 0.387$.  
Since the singularity of interest is about factor $2.1$ closer than the next nearest, 
any singularity at $\, w\, =\, 1/2$  will be observable in an $N$ term Euler transformed 
$z$ series with corrections of order $\, (2.1)^{-N}$.  This factor $\, (2.1)^{-N}$ is also 
the bound we can put on any $\, w\, =\, 1/2$  singularity amplitude if a ratio test 
of coefficients does not indicate a singularity at $\, z\, \approx \, -0.184$.
We find the absence of such a singularity for the $A(z)$ series which we have generated
to length $\,N=\, 1100$.  Because the $\, A(z)$ series multiplies $\, \ln(z)$, this also 
indicates that any possible singularity at $\, w=\, 1/2$ will have an amplitude 
independent of the index $\, m$ specifying the logarithmic branch of
the $z=0$ ($w\, =\,(3 -\sqrt{5})/2$) singularity.

We find that for $n =\, \pm 1$, the $\, B(z)$ series is also not singular at 
$\, z\, \approx \, -0.184$.  For other $n$ values a singularity is clearly indicated 
and by a coefficient ratio analysis of different $\, n$ series completely analogous 
to what was done for the $\, A_1$ and $\, A_2$ series in (\ref{eq:A1},\ref{eq:A2}) 
we find that the singularities at $\, w=\, 1/2$ have amplitudes proportional to the 
branch dependent multiplicity factor $\,(n^2-1)^2$.  A more detailed analysis 
involving explicit fitting of the $\, B(z)$  series coefficients yields the singularity 
amplitude which we have confirmed by direct matching of the $\, \tilde{\chi}^{(5)}$ 
series in $z$ to ODE solution series about $\, w=\, 1/2$.  The result 
for the singular part of $\tilde{\chi}^{(5)}$ at $\, w=\, 1/2$ is
\begin{eqnarray}
\label{finalresult}
&&	\tilde{\chi}^{(5)}_{\rm sing}\,  =\, \, 
 \big[(n^2-1)^2 (2\sqrt{6})/(315\, \pi)\big]\cdot\, (1-2\, w)^{7/2}
 \nonumber \\
&& \times   \, \, [1\, +41\, (1\, -2w )/12\, +26557\, (1-2w)^2/3168\, +\, \cdots ], 
\end{eqnarray}
where the amplitude has been verified to our numerical accuracy of 
about 250 digits.  We have not identified a Landau singularity associated
with (\ref{finalresult}). Finding this Landau singularity in the $\, {\tilde \chi}^{(5)}$
integrand remains as the major unsolved challenge of this paper.

\vskip .1cm

\section{Conclusion}
\label{concl}
\vskip .1cm

We have completed the quest begun in~\cite{experim} for the exact integer
arithmetic ODE satisfied by  $ \tilde{\chi}^{(5)}$.  While most explicit results are 
far too extensive to be published here, a selection can be found on the 
website~\cite{http}.  These include $\, L_{24}$ and $E^{(5)}$ defined in 
(\ref{eq:L24})--(\ref{eq:E5}), the explicit factorization 
$\, L_{24} = L_{12}^{(\rm left)}\cdot L_{12}^{(\rm right)}$,
the operator $L_{29}$ and the exact integer high temperature series
for  $ \tilde{\chi}^{(5)}$ to 8000 terms generated from (\ref{eq:L24})--(\ref{eq:E5})
and numerical coefficients to 800 digits for the $ \tilde{\chi}^{(5)}$ series at the 
ferromagnetic point to supplement (\ref{A4}).

We have used the exact ODE to resolve at least one issue that was left 
undecided in~\cite{High}.  In particular, we have shown in section~\ref{proof} 
that the operator $\, L_{12}^{(\rm left)}$ cannot be factored.  The techniques 
we have described, both for factorization and proving the converse, are not 
all well known and we expect they will find application in other problems.
In addition we have shown that  $L_{12}^{\rm (left)}$ isn't a symmetric $n'$th 
power of a lower order operator nor is it a symmetric product of two (or more) 
operators  if the order of the symmetric product reaches its maximum value.

An issue raised in~\cite{experim}  has only been partially resolved.  
No obvious candidate for a Landau singularity in the integral representation 
of $ \tilde{\chi}^{(5)}$  at $\, w=\, 1/2$  could be found and because a Landau 
integrand singularity is a necessary condition\footnote{Hwa and 
Teplitz (see page 6 in \cite{Hwa}) give \cite{Hadamard} 
and \cite{Polk} as the original sources for 
the necessary requirement.}  \cite{Hwa} for an integral to be singular, 
it was conjectured that $ \tilde{\chi}^{(5)}$, necessarily defined by analytic 
continuation from its series representation, would not be singular at 
$w=1/2$\footnote{The abstract in~\cite{experim} erroneously stated the 
absence of a singularity without caveat.  In view of the findings of this paper 
the conclusion that $ \tilde{\chi}^{(6)}$ is singularity free at $w^2=1/8$ must 
now also be discounted.}.  We have now, in section~4, shown that 
$ \tilde{\chi}^{(5)}$ is singular at $w=1/2$ and so clearly have identified all 
the singularities of $ \tilde{\chi}^{(5)}$ by an analysis of the ODE it satisfies.  
The $w=1/2$ singularity in $ \tilde{\chi}^{(5)}$ proves the existence of an 
associated Landau integrand singularity, but we have made no progress in 
identifying this Landau singularity.  And so our goal of unifying the Landau 
integrand analysis with the ODE approach to the behaviour of the general 
$ \tilde{\chi}^{(n)}$ continues to elude us.

\ack
IJ and AJG are supported by the Australian Research Council. 
The calculations  would not have been possible without a generous grant 
from the National Computational Infrastructure (NCI) whose National Facility 
provides the national peak computing facility for Australian researchers.
This work has been performed without any support
 from the ANR, the ERC, the MAE. 

\appendix

\section{The exact arithmetic $L_{12}^{(\rm right)} = \,  L_1 \cdot L_{11}$}
\label{exactL1tilde}
\vskip .1cm

It was reported in~\cite{High} that $\, L_{12}^{(\rm right)}\, = \,\, L_1 \cdot L_{11}$
with a solution $S_{12}$ that satisfies $\,L_{11}(S_{12}) =\,  P$, $\,\,\,\,L_1(P) =\, 0$ 
where  $\, P$ is a polynomial. The order eleven linear differential operator 
$\, L_{11}$ has a direct-sum decomposition
\begin{eqnarray}
\label{direct}
L_{11} \,=\,\,\, \left( Z_2 \cdot N_1 \right) \oplus V_2 \oplus 
\left( F_3 \cdot F_2 \cdot L_1^s \right).
\end{eqnarray}
where $\, Z_2$ is a second order operator also occurring in the factorization of 
the linear differential operator associated with $ \, \tilde{\chi}^{(3)}$,
$\, V_2$ is a second order operator equivalent to the second order operator  
associated with  $ \, \tilde{\chi}^{(2)}$ and  $\, F_3$ and $\, F_2$ are known
in exact arithmetic~\cite{High}. The exact $\, L_{11}$  was obtained in~\cite{High} 
from (\ref{direct}) and is given\footnote[5]{Note that $\, L_{11}$ in~\cite{http} 
is given in monic form. Here $\, L_{11}$ is understood to be normalized to a 
sum of products of polynomials times powers of $\, w  \cdot \rmd/\rmd w$.} 
in~\cite{http}. Our goal here is to determine the exact $\, L_{12}^{(\rm right)}$ or 
equivalently the exact  solution $\,S_{12}$ which gives the polynomial
 $\, P =\,  L_{11}(S_{12})$ and then $\, L_1$.  
 
One possible form for this solution is
\begin{eqnarray}
\label{eq:S12}
&&S_{12}\,\, = \, \, \,\, 
\, \ln(w)\cdot S_9\,  -2w^3\, -6w^4-136\, w^5/3 \, -3161\, w^6/24\, \nonumber \\
&&\qquad  -3073\, w^7/4\, -245147\, w^8/120 \,  -334031\, w^9/30\, \nonumber \\
&&\qquad -872442\, w^{10}/35\,-19192553\, w^{11}/140\, +\, \, O(w^{13})
\end{eqnarray}
where
\begin{eqnarray}
\label{eq:S9}
&&S_9 =\,  w^2\, +4\, w^3\, +30\, w^4\,  +120w^5\, +690 \, w^6\,
          +2760\, w^7\, +14280\, w^8\,\nonumber \\
&&\qquad +57120\, w^9\, 
	        +279090\, w^{10}\, +1116360\, w^{11}\,
 \nonumber \\
&&\qquad +5261244\, w^{12}\,  +\, \, O(w^{13})
\end{eqnarray}
is annihilated by the order nine operator 
$V_2 \oplus  N_1 \oplus \left( F_3 \cdot F_2 \cdot L_1^{s} \right) $. 
The rational coefficients in (\ref{eq:S12}) 
are sufficiently small that the four modulo prime solutions of (\ref{eq:L24})
reported in~\cite{High} will yield $\, S_{12}$ exactly to $\, O(w^{19})$.  This
in turn means that $\,P\, = \,\, L_{11}(S_{12})$ modulo any prime is given exactly 
to $O(w^{19})$.  We now observe that the degree of $\, P$ depends on the 
$\, L_{11}$ representation and is 18 when either the minimum degree or
minimum degree plus one representation of $L_{11}$ is used.  This implies 
that, with this degree restriction on $L_{11}$ and without any further
input from $\, \Phi^{(5)}$, the equation $P_{18} \,=\,\,  L_{11}(S_{12})$
determines $P_{18}$ modulo any prime exactly.  The same equation can 
then be used in recursion mode to extend the known $\, S_{12}$ modulo a
prime series of $O(w^{19})$ to any desired length.  Clearly these steps can 
be repeated for as many primes as necessary to get the $S_{12}$ coefficients
in exact arithmetic by rational reconstruction.  Then, once $S_{12}$ has been 
extended to $O(w^{137})$, one can switch back to the minimum order 
representation of $\, L_{11}$ to get $P_{136}\,  =\,\,   L_{11}(S_{12})$, followed 
by $L_{1}$ and the minimum order $\, L_{12}^{(\rm right)}$ in exact arithmetic.  
Note that we make no attempt to reconstruct $\, P_{18}$ in exact arithmetic.  
This polynomial is of no particular use and since it is generated from a 
non-minimal order $L_{11}$ we expect its coefficients to be equally 
outrageously large.

The polynomial $P_{136}$ is proportional to
\begin{eqnarray}
\label{P136}
\fl \qquad \qquad 5544\, w^5\, +197765\, w^6\, +419883469\, w^7/70\, -24564638026\, w^8/35 
\, +\, \cdots \nonumber \\
\fl \qquad \qquad \qquad -2^{171}\cdot 134196985\, w^{135}/363\,
 -2^{173}\cdot 80400525\, w^{136}/847.
\end{eqnarray}
and the operator $L_1$ that annihilates it is obviously proportional to 
$P_{136}{\rmd \over \rmd w} -{\rmd P_{136}\over \rmd w}$.  Straightforward 
operator multiplication $L_1\cdot L_{11}$ yields an $L_{12}^{\rm (right)}$ 
and this can be normalized to monic or non-monic form as desired.  Note 
that $P_{136}/w^5$ becomes the apparent singularity part of the head 
polynomial of $L_{12}^{\rm (right)}$.  The corresponding apparent polynomial 
$P_{\rm app}$ of $L_{11}$ divides out of the product $L_1\cdot L_{11}$ 
as a common factor; this serves as a useful check of our algebra.  
The minimum order $L_{12}^{\rm (right)}$ we have obtained 
can be found in \cite{http}.

The fact that the series $S_9$ multiplying the logarithm in (\ref{eq:S12}) 
is annihilated by an operator of order $M = 9 < 11$ suggests that it might 
be possible to ``push" either $V_2$ or $Z_2$ to the left of $L_1$.  And 
indeed we found that a $U_1$, equivalent to $L_1$, occurs at the left of the 
$L_9 = V_2 \oplus  N_1 \oplus \left( F_3 \cdot F_2 \cdot L_1^{s} \right) $ that 
annihilates $S_9$ in (\ref{eq:S9}).  Next, our aim is to see whether 
$U_1$ can be pushed further to the right.  This amounts to considering the 
various factorizations of $L_9$; for instance the factorization where the 
equivalent of $V_2$ occurs at the leftmost position and there exists a direct sum 
of this with $U_1$.  There is such a direct sum showing that an equivalent 
to $U_1$ occurs at the left of $N_1 \oplus \left( F_3 \cdot F_2 \cdot L_1^{s} \right) $.  
The process continues until it is no longer possible to find a direct sum of 
our equivalent to $L_1$ and any one of the right factors.  This final factorization has a solution
\begin{eqnarray}\label{eq:S5}
	S_5 = \ln(w)\cdot w^2\cdot \left[1+\sqrt{(1-4\,w)/(1+4\,w)}\right] \Big/(1-4\,w)^2  \nonumber \\ 
	\qquad   + 64\,w^5/9+200\,w^6/9+712\,w^7/3+10708\,w^8/15 \nonumber \\ 
   \qquad +253888\,w^9/45+1771852\,w^{10}/105+12245672\,w^{11}/105 \nonumber \\
   \qquad +331735612\,w^{12}/945  +O(w^{13}) 
\end{eqnarray}
which can replace $S_{12}$ from (\ref{eq:S12}) as one of the 12 linearly 
independent solutions of $L_{12}^{\rm (right)}$.  The operator that annihilates $S_5$ is
\begin{equation}\label{eq:L5t}
	U_5 = \tilde{L}_1 \cdot (N_1 \oplus (F_2\cdot L_1^s))	
\end{equation}
where if we define $U_4 = N_1 \oplus (F_2\cdot L_1^s)$ to be given in our standard 
non-monic form then $U_4(S_5) \propto P_{19}\, w^2/(1-4w)$ with

\begin{eqnarray}
\fl \quad	P_{19} = 9+36\, w+18\, w^2-2064\, w^3+4581\, w^4+59584\, w^5-143476\, w^6 -898464\, w^7 \nonumber \\
 \fl \quad \qquad      +124724\, w^8+813120\, w^9+9220240\, w^{10}+55704896\, w^{11}+65556224\, w^{12}  \nonumber \\
 \fl \quad \qquad        -253883392\, w^{13}-406194176\, w^{14}+1318182912\, w^{15}+2053013504\, w^{16}  \nonumber \\
 \fl \quad \qquad     +368443392\, w^{17}-454033408\, w^{18}-272629760\, w^{19}.
\end{eqnarray}
The operator  that annihilates $U_4(S_5)$ is 
\begin{equation}
\fl \qquad \qquad
\tilde{L}_1=\Big[\, w(1-4\, w)P_{19}\Big]{\rmd \over \rmd w} -\left[(2-4\, w)P_{19}+w(1-4\, w){\rmd P_{19} \over \rmd w}\right]
\end{equation}
and again it is straightforward to combine this $\tilde{L}_1$ with the other exact arithmetic 
operators in (\ref{eq:L5t}) to get $U_5$.

An incidental remark on the solution (\ref{eq:S5}) is that this is one of the few 
examples of a (partial) analytic solution of $L_{12}^{\rm (right)}$.  Other simple solutions are 
$w^2/(1-4\,w)^2$, $w^2/(1-4\,w)/\sqrt{1-16\,w^2}$ and $w^2\,(1+4\,w)\, _2F_1(3/2,3/2;1;16\,w^2)$, 
$w^2\,(1+4\,w)\,_2F_1(3/2,3/2;3;1-16\,w^2)$ 
associated with $L_1^s$, $N_1$ and $V_2$, respectively.  A more complicated case is that of 
$Z_2$ for which we refer\footnote[5]{Note a misprint in Appendix C in \cite{High}, in eqn (C7). 
The right-hand side should in fact be replaced by $8^2(1-2w)^2(1+2w)^2/(1-w)^2/(1-4w)^2.$}  
the reader to \cite{High}. 

In conclusion, every solution of $L_{12}^{\rm (right)}$ can now 
be obtained as a linear combination of solutions of operators of order $M \le 6.$

\vskip .1cm

\section{$\, {\tilde \chi}^{(5)}$ at the ferromagnetic point}
\label{ferro}
\vskip .1cm
To find $\, \Phi^{(5)}$ in the vicinity of the point 
$y=\, 0$ where $\, 2y \, =\, 1\, -4w$ requires that we first 
find a solution to the inhomogeneous equation (\ref{eq:L24})
 near $y=\, 0$.  To this end we develop 
the elliptic integrals in (\ref{eq:E5}) as series in $\, y$
 after which finding a particular integral
 by series recursion is straightforward.  What remains 
is then the matching of the $\, \Phi^{(5)}$
 series about $\, w=\,0$ to this particular integral
 plus the 24 homogeneous solution series $S_i$ 
satisfying $\, L_{24}(S_i) =\, 0$.  This too is a 
straightforward exercise although clearly one must
 pay attention to potential errors arising both
 from numerical round-off and series
 length truncation.  A useful remark 
in this regard is that it is very advantageous
 to work with $\Phi^{(5)}$ series in $s$ rather than 
$ w =\, s/2/(1+s^2)$.  This arises because of the
 non-linear relationship $\, 1\, -4w\,   \approx \, (1-s)^2/2$
  at the ferromagnetic critical point.

To understand the effect of the choice of series variable we note
that in general the error resulting
from truncation of a series in $x$ at $ N$ terms when
evaluated at $ x_m$ scales roughly as $(x_m/x_s)^N$
where $\, x_s$ is the distance to the nearest singularity
from\footnote[1]{We assume for purposes
of our qualitative discussion here that all functions
are approximately of unit magnitude.} $x =\,0$.
As an illustration, from $\, \Phi^{(5)}$  series of $N=\, N_w$
terms and a matching point at $\, w_m \, = \, (1\, -2y_m)/4$, the
series truncation error is $\, \approx \, (4\, w_m)^N $
$= \, (1-2\,y_m)^N \, \approx \, \exp(-2\, N_w\, y_m)$
for small $y_m$.  Now suppose
the solution series $S_i$ have been evaluated to $N=\, N_y$ 
terms.  The solutions with the smallest radius of
convergence\footnote[2]{This $ w_s$ is the same as 
that in the discussion leading to and following (\ref{eq:A1},\ref{eq:A2}).},
$y_s =\, 1/2-2\, w_s \, \approx \,  0.183$, 
determine the truncation error which is 
$ \approx \,  (y_m/0.183)^N \, \approx \,\,  \exp(-N_y\, \ln(0.183/y_m))$. 
The optimal choice of matching point is where the errors are roughly equal, i.e.
\begin{eqnarray}
\label{A1}
2\,  N_w\cdot y_m\,\, = \,\,\, N_y \cdot \ln(0.183/y_m).
\end{eqnarray}
With $\,N_w\,=\,8000$ and $\, N_y\,=\,800$, 
condition\footnote[5]{We have reached
 such values for exact rational coefficient
 series without too much computing effort.}  (\ref{A1}) yields 
$y_m =\,0.0577$ with a resulting 
truncation error $\,\approx \, 10^{-400}$.  This estimate of 400 digit
accuracy is overly optimistic since
the $\, \Phi^{(5)}$  matching requires the evaluation of 23
derivatives which invalidates our assumption of functions all of unit magnitude.

Consider now what happens if the matching is done using  $\, \Phi^{(5)}$  as a
series in $\, s$ of $N =\, N_s=N_w$ terms.  The truncation error is
$\approx \, (s_m)^N  \approx\,  (1-2\, \sqrt{y_m})^N \,$
$ \approx \, \exp(-2\, N_w \, \sqrt{y_m})$
for small $y_m$.  The match point choice that replaces (\ref{A1}) is
\begin{eqnarray}
\label{A2}
2\,  N_w \cdot  \sqrt{y_m} \, \, = \,\, \, N_y \cdot  \ln(0.183/y_m), 
\end{eqnarray}
and with the same series lengths as above, $y_m\, =\,\, 0.01535$ 
with resulting truncation error $\,\approx \,  10^{-860}$. 
 Again this must be overly optimistic but it does illustrate 
the dramatic improvement achieved
 with no extra computational cost.

Our results for the behaviour of $\, \Phi^{(5)}$  at the ferromagnetic point are as 
follows:  We use throughout the definition $\,2\,y\,=\,\, 1\, -4w$ as above.  Then, to
$\,O(y^{10})$, the 5-particle contribution is
\begin{eqnarray}
\label{A3}
\fl \qquad 120\, \pi^4 \cdot \Phi^{(5)} = \, 
({\tilde \chi}^{(1)} \, -60\, {\tilde \chi}^{(3)} \, +120\, {\tilde \chi}^{(5)}) \cdot  \pi^4 
 \, = \, \nonumber \\
\fl \qquad  \ln^4(y/4)\cdot (-5/128\, -25y/512\, -75y^2/512\, +6455y^3/8192\,
\nonumber \\
\fl \qquad \qquad  +42305y^4/16384 \, +49935y^5/8192\, +444955y^6/32768
\nonumber \\
\fl \qquad \qquad +64409075\, y^7/2097152\, +306977235\, y^8/4194304
\nonumber \\
\fl \qquad \qquad +3131527805y^9/16777216) \nonumber \\
\fl \qquad +\ln^3(y/4)\cdot (-235/192\, -1043\, y/512\, -559\, y^2/256\,
\nonumber \\
\fl \qquad \qquad  +1007843\, y^3/1032192
	        +265627475\, y^4/22708224\, \nonumber \\
\fl \qquad \qquad +21838013489\, y^5/590413824\, 
	        +57270165499\, y^6/590413824\, \nonumber \\
\fl \qquad \qquad +315303684751873\, y^7/1284740481024\,\nonumber \\
\fl \qquad \qquad  +31001275613111851\, y^8/48820138278912\, \nonumber \\
\fl \qquad \qquad +336259319399213305\, y^9/195280553115648)
\nonumber \\
\fl \qquad +\ln^2(y/4)\cdot (-1225/192\, +306283\, y/15360\, -859553\, y^2/215040\,
 \nonumber \\
\fl \qquad \qquad -20921323001\, y^3/433520640\, -10704883015027\, y^4/104911994880
\nonumber \\
\fl \qquad \qquad -12546069917407919\, y^5/70920508538880\, \nonumber \\
\fl \qquad \qquad -1021232328273251\, y^6/3546025426944\, \nonumber \\
\fl \qquad \qquad -1219521777522960411013\, y^7/2623491451870248960
\, \nonumber \\
\fl \qquad \qquad -1428705274437356046530557\, y^8/1894160828250319749120\,
 \nonumber \\
\fl \qquad \qquad -3094013953756589876579173\, y^9/2525547771000426332160)
\nonumber \\
\fl \qquad     +\ln(y/4)\cdot \left(\sum_{k=0}\,  C_1(k)\cdot  y^k \right)
 \quad +\sum_{k=-1}\,  C_0(k)\cdot y^k 
\end{eqnarray}
where the constant arrays $C_1$ and $C_0$ are only known as 
floating point values.  Truncated values are
\begin{eqnarray}
\fl \qquad C_1(0) = -24.57942277608500580980766691884235675672213367715281
\, \cdots \nonumber \\
\fl \qquad C_1(1) = 100.33228026112198073984483757868964242744468431106263
\, \cdots \nonumber \\
\fl \qquad C_1(2) = 153.62912225095690287937642813703191769479633799429441
\, \cdots \nonumber \\
\fl \qquad C_1(3) = 96.268269775450223433437918344019644004451927310385706
\, \cdots \nonumber \\
\fl \qquad C_1(4) = -93.06087224694711987327189827852646701359704351289178
\, \cdots \nonumber \\
\fl \qquad C_1(5) = -520.0604636206711421998793353361815107994358750212855
\, \cdots \nonumber \\
\fl \qquad C_1(6) = -1469.482403857530493377284131808565295376128516174659
\, \cdots \nonumber \\
\fl \qquad C_1(7) = -3700.403260075047090620399746433870069155277603230413
\, \cdots \nonumber \\
\fl \qquad C_1(8) = -9303.076094072409117026994688305643439126188417678639
\, \cdots \nonumber\\
\fl \qquad C_1(9) = -24403.45115956242143888235859126942238116466146718368
\, \cdots \nonumber
\end{eqnarray}
\begin{eqnarray}
\label{A4}
\fl \qquad C_0(-1)= 23.164561203366712117448548909598809004328248610670601
\, \cdots \nonumber \\
\fl \qquad C_0(0) = -117.5551740623092343089105578149581427399163833071621
\, \cdots \nonumber \\
\fl \qquad C_0(1) = 256.41151149949623257928100314293461191887809056234918
\, \cdots \nonumber \\
\fl \qquad C_0(2) = 350.91585906790101529219839116619800691812963977611999
\, \cdots \nonumber \\
\fl \qquad C_0(3) = 293.47893916768186306623236360668073915143408501485741
\, \cdots \nonumber \\
\fl \qquad C_0(4) = 109.38934363323783704769948377240709436366681466990651
\, \cdots \nonumber \\
\fl \qquad C_0(5) = -373.9288055773641613398692905165600188023113404992740
\, \cdots \nonumber \\
\fl \qquad C_0(6) = -1561.022604360024268401667883538021798990658115951832 
\, \cdots \nonumber \\
\fl \qquad C_0(7) = -4574.248865947314517242004568453306010341709549402520
\, \cdots \nonumber \\
\fl \qquad C_0(8) = -12595.46119191674598711644766973884064574672223482538
 \, \cdots \nonumber \\
\fl \qquad C_0(9) = -35224.55961370779008532569450932078801945477728287548
\, \cdots \nonumber \\
\end{eqnarray}
while values to about 800 digits can be found in~\cite{http}.
The 3-particle function, to $O(y^{10})$, is  
\begin{eqnarray}
\label{A5}
\fl \qquad -6\, \pi^2 \cdot \Phi^{(3)}\,\,  = \, \,\,
 (\tilde{\chi}^{(1)}\, -6\, \tilde{\chi}^{(3)})\cdot \pi^2 \,\,  = \, \,\,
 \,\,\nonumber \\
\fl \qquad  \ln^2(y/4)\cdot 
\big(3/32\, -3y/128\, +9\, y^2/128\, +87\, y^3/2048\,\nonumber \\
\fl \qquad \qquad  -39\, y^4/4096\, -423\, y^5/4096
	        -555\, y^6/2048\, -301647\, y^7/524288
\,\nonumber \\
\fl \qquad \qquad -1185363\, y^8/1048576
	        -8996817\, y^9/4194304\big)
 \nonumber \\
\fl \qquad +\ln(y/4)\cdot  \big(23/32\, +227\, y/1280\, -2047\, y^2/4480\, -88949\, y^3/122880
\nonumber \\
\fl \qquad \qquad  -20562503\, y^4/18923520\, -810591833\, y^5/492011520
\nonumber \\
\fl \qquad \qquad -316471567\, y^6/123002880\, -4445362809179\, y^7/1070617067520
\nonumber \\
\fl \qquad \qquad  -3085016829083\, y^8/447070863360\nonumber \\
\fl \qquad \qquad -273049228448281\, y^9/23247684894720\big)\nonumber \\
\fl \qquad +\big(41/96\, -21169\, y/38400\, -225583\, y^2/3763200\,
 +2621231\, y^3/154828800\, 
\nonumber \\
\fl \qquad \qquad +15533081173\, y^4/262279987200\, \nonumber \\
\fl \qquad \qquad +151705656477979\, y^5/1595711442124800
\, \nonumber \\
\fl \qquad \qquad +30692556260057\, y^6/265951907020800
\, \nonumber \\
\fl \qquad \qquad +1747927197871890533\, y^7/19676185889026867200
\, \nonumber \\
\fl \qquad \qquad -41073304786882831381\, y^8/655671055932802990080
\, \nonumber \\
\fl \qquad \qquad -776816275820131600824097\, y^9/1534270270882758996787200\big)
 \nonumber \\
\fl \qquad +\big(9 \, \psi^{(1)}(1/3)/8\, -3/2\, -3\, \pi^2/4\big)\cdot \big(1/y-3/2\, -5y/24\, -3y^2/8\,
\nonumber \\
\fl \qquad \qquad -1801\, y^3/3456
	        -1649\, y^4/2304\,-187999\,y^5/186624
\,\nonumber \\
\fl \qquad \qquad -45617\,y^6/31104\,
	        -17592665y^7/7962624\,\nonumber \\
\fl \qquad \qquad -164030851\,y^8/47775744\,
	        -9425604977y^9/1719926784\big)
\end{eqnarray}
where $ \psi^{(1)}(x)=\,   \psi'(x) $ is the polygamma function 
($\psi(x)=\, \Gamma'(x)/\Gamma(x)$).  The coefficient of $\,\, 1/y$ in (\ref{A5})
was determined by Tracy~\cite{Tracy} while the 
higher order terms are recursively generated
from the ODE satisfied~\cite{ze-bo-ha-ma-05c} by $ \Phi^{(3)}$.  Finally,
\begin{eqnarray}
\Phi^{(1)} \,  =  \, \, \tilde{\chi}^{(1)}\, \, \,
  = \, \, \,  1/(4y) \, -1/2.
\end{eqnarray}
When these results are combined and expressed
 in terms of $ \tau \, =\, (1/s\, -s)/2$,
i.e. $\,2\, y\, =\, \tau^2/(1+\tau ^2+\sqrt{1+\tau^2})$,
 one confirms the low order result
 for $\, \chi^{(5)}$ in equation (48) of~\cite{experim}.  
We also confirm the 500 digit expression for the 
$\, \chi^{(5)}$  amplitude $\, D_5$ given
by Bailey \etal~\cite{Bailey}.  The explicit connection is
\begin{eqnarray}
D_5 \,=\,\, \,\,   52\, \pi^4/5\, -12\, \pi^2 \, \psi^{(1)}(1/3)\, \, 
+16\, \pi^2\, +16\, C_0(-1)/15
\end{eqnarray}
with $\, C_0(-1)$ given in (\ref{A4}).

The 5-particle contribution (\ref{A3}) to $O(y^{10})$
 can be extended to arbitrary order 
by a purely local analysis of the ODE (\ref{eq:L24}).  This 
makes (\ref{A3}) particularly useful for
 those problems in which one wants to analytically 
continue $\, \Phi^{(5)}$ beyond the ferromagnetic 
singularity.  Of the constants in (\ref{A4}), only
 the fifteen $\, C_1(0..5)$ and $\, C_0(-1..7)$ depend
 directly on solution matching.  All higher order 
coefficients are fixed by recursion
 relations with rational coefficients.  The rationals are quite large, even
 in the simplest example which reads
 \begin{eqnarray}
\label{BB8}
\fl \qquad 28235138014521201151215868809287621440 \cdot  C_1(6)
 \, \, = \, \,  \, \,-\, {{n_1} \over {d_1}}\nonumber \\
\fl \qquad	  -37737948182868464640476571846064977272634973/1168128 \cdot  C_1(0)
\nonumber \\
\fl \qquad	  -2562052253560544220030628882123379039659169/86528 \cdot   C_1(1)
\nonumber \\
\label{A8}
\fl \qquad	  +8187149177658628591337051452109148482747/312 \cdot C_1(2)
\nonumber \\
\fl \qquad	  +158617170239026490613752192866932708861763/1872\cdot  C_1(3)
\nonumber \\
\fl \qquad  -140695455847623820725386797097484481682383/702 \cdot  C_1(4)
\nonumber \\
\fl \qquad	  +133960461264493957765301921009624203227 \cdot  C_1(5),
\end{eqnarray}
with
\begin{eqnarray}
\fl \qquad
n_1 \, = \, \,\ 15256931516199856571741494453893800774223260951365779571949, \nonumber \\
\fl \qquad
d_1\, = \, \, 31365304106405068800. \nonumber 
\end{eqnarray}
Floating point recursion in $\,y$ is unstable to the extent that relative errors increase 
by about factor 40 per order in the $\,y$ series.  If one wants, say, 1500 terms, 
one must start with 2400 digits more than the desired final accuracy. 
One must also use directly as starting values only the coefficients
$\,C_1(0..5)$ and $\,C_0(-1..7)$ from (\ref{A4}). All other coefficients are to be 
obtained to the necessary higher accuracy directly from the ODE (\ref{eq:L24}) 
locally around $y=\, 0$, now treating the starting values $C_1(0..5)$ 
and $C_0(-1..7)$ as if exact.

\begin{figure}
\includegraphics[width=12cm]{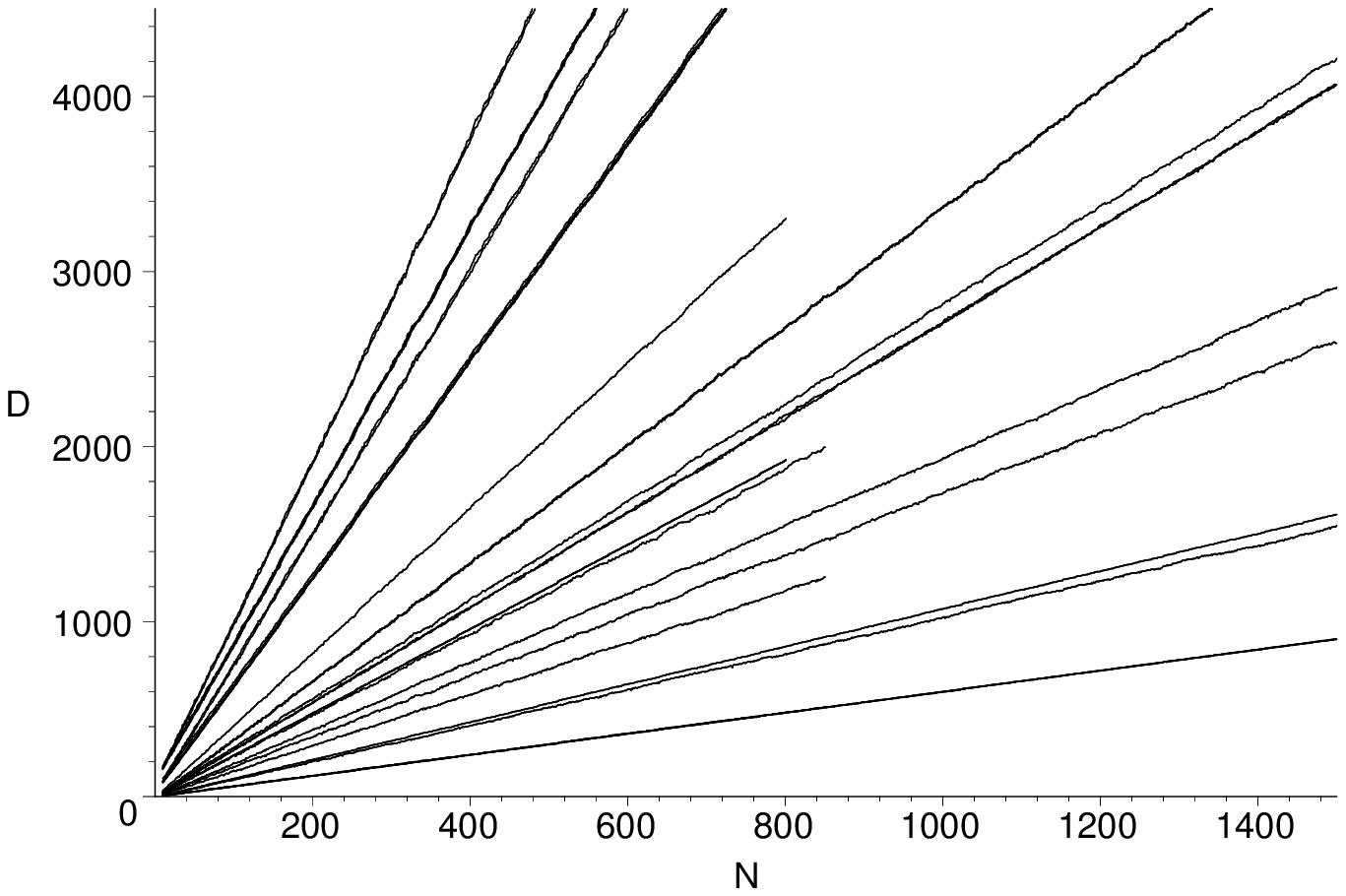}
\caption{\label{fig:denom}
The number of digits $D$ in the denominators of the rational coefficients $c_N$ 
in series $y^p \ln^q(y/4) \sum c_N y^N$ that arise in ODE solutions around $2y = 1-4w = 0$.}
\end{figure}

The problem of numerical instability in series generation around rational points 
can be avoided by generating all series in exact arithmetic.  In that case the series 
coefficients are in general rational and our observation, admittedly rather limited, 
is that the growth of the denominators is at most exponential.  An example of 
denominator growth is shown in Fig.~\ref{fig:denom} for all series in $y$ that we 
have generated for doing the ferromagnetic point series matching described above.  
A growth rate faster than exponential would make exact arithmetic series generation 
impractical so it is of some importance to know whether the observed exponential 
growth is a general feature.  It is reminiscent of G-series \cite{Dwork} but we 
are unaware of any theorem that guarantees this behaviour for expansions of 
integrals with algebraic integrands about singular points.

\section{Right division operators for irrational singular points}
\label{rightdiv}
\vskip .1cm

We wish to determine a right division of a minimum order $L_N(w)$ from 
solutions identified with the non-rational zeros of a factor $h(w)$ of 
the head polynomial of $\, L_N(w)$.  Examples of $h(w)$ treated in the 
text are $\, 1\, +3\, w\, +4\, w^2$ and the case 3 and 4 polynomials from 
Appendix~C in~\cite{experim}.  If such a polynomial has an ``accidental'' 
modulo prime $\, p$ factorization $ h(w)=\, (w\, -w_p)\, h_p(w)$, we 
can still generate series solutions about $\,x =\, 0$ where $x =\, w-w_p$.  
Furthermore, suppose the solution $S(x)$ has been tested and is found 
to be annihilated by a minimum order $M<\, N$ operator.  If the code used 
for the testing is that described in section~3 of~\cite{experim} the result 
will be of the form
\begin{eqnarray}
\label{B1}
L_M(x)\,  =\, \,\,  \,\, 
\sum_{n=0}^{M} \,  f_n(x) \cdot
 \Bigl(x \cdot {{\rmd} \over {\rmd x}}\Bigr)^{M-n}
\end{eqnarray}
where the head polynomial $\, f_0(x)$ is normalized such that $f_0(x=0)=\, 1$.  
Note that $ w_p$ is not the modulo $\, p$ representation of a root of $ h(w)$ 
and both it and $x$ have no meaning independent of the specific modulo $\, p$ 
factorization $\, h(w)\,=\,(w\, -w_p)\, h_p(w)$.  Consequently there does not exist 
any exact ``$L_M(x)$'' to be reconstructed from multiple $L_M(x)$ generated 
using different primes.  Replacing $x$ by $w\, -w_p$ in (\ref{B1}) generates 
an $\, L_M(w)$ but only when this is correctly normalized to remove all 
dependence on the ``accidental'' $ w_p$ can it be used for a reconstruction 
of an exact $L_M(w)$.  The steps for doing this are as follows.

To begin, note that (\ref{B1}) can be expressed as a sum of polynomials times 
powers of $\rmd/\rmd x$. The head polynomial in this representation is clearly 
$x^M \, f_0(x)$.  The modulo $\, p$ factorization of this will contain as factors 
$\, x^M \, [h_p(x+w_p)]^K$, typically with $K \, <\, M$.  These factors can be 
written as $x^{M-K}\, [x\, h_p(x+w_p)]^K \, =\,  x^{M-K}\, [h(x+w_p)]^K \, =\,  
(w-w_p)^{M-K}[h(w)]^K$.  Dependence on the ``accidental'' $w_p$ appears 
only in the factor $x^{M-K}$ and this can be divided out.  So too can the 
$ w=\, 0$  normalization constant $C=\, x^K \, f_0(x)$ evaluated at $x=\, -w_p$. 
 The result is the right division (modulo $p$) operator which can be cast 
into the same form as (\ref{B1}), namely
\begin{eqnarray}
\label{B2}
&&L_M(w) =\,\,  \left. \frac{L_M(x)}{x^{M-K}C}\right|_{x=\, w\, -\, w_p } \quad
 {\rm modulo} \, \,\, p\nonumber \\
&& \qquad =\, 
\sum_{n=0}^M \, g_n(w) \cdot
 \Bigl(w \cdot {{\rmd} \over {\rmd w}}\Bigr)^{M  \,- \, n}
\end{eqnarray}
where each $ \,g_n(w)$ is a polynomial and, by construction, $g_0(w=0)=\, 1$.  
The exact $L_M(w)$ can now be reconstructed from different prime based 
versions of (\ref{B2}).

\end{document}